\documentclass[a4paper, 11pt, oneside]{article}
\pdfoutput=1
\usepackage{jheppub}
\usepackage{graphicx,wrapfig,float,slashed,indentfirst}
\usepackage{amsmath,amssymb,epsfig,xcolor}
\usepackage{fancyhdr}
\usepackage{epstopdf}
\usepackage{graphics}
\usepackage[mathscr]{eucal}
\usepackage{feynmp}
\usepackage{xspace}
\usepackage{listings}
\usepackage{gensymb}
\usepackage{pstricks}
\usepackage{xspace}
\usepackage{braket}
\usepackage{subfigure}
\usepackage{array}
\usepackage{feynmp}
\newcommand{\SARAH}{{\tt SARAH}\xspace}
\newcommand{\PYRATE}{{\tt Pyrate}\xspace}
\newcommand{\SPheno}{{\tt SPheno}\xspace}
\newcommand{\MICRO}{{\tt micrOMEGAs}\xspace}

\newcommand{\te}{\renewcommand{\arraystretch}{1.4}}

\newcommand{\nc}{\newcommand}
\nc{\beq}{\begin{equation}}  \nc{\eeq}{\end{equation}}
\nc{\bea}{\begin{eqnarray}}  \nc{\eea}{\end{eqnarray}}
\nc{\baa}{\begin{array}}     \nc{\eaa}{\end{array}}
\nc{\bit}{\begin{itemize}}   \nc{\eit}{\end{itemize}}
\nc{\ben}{\begin{enumerate}} \nc{\een}{\end{enumerate}}
\nc{\bce}{\begin{center}}    \nc{\ece}{\end{center}}
\nc{\bpm}{\begin{pmatrix}}   \nc{\epm}{\end{pmatrix}}
\nc{\bvt}{\begin{verbatim}}  \nc{\evt}{\end{verbatim}}
\nc{\non}{\nonumber} 
\newcolumntype{M}{>{$\vcenter\bgroup\hbox\bgroup}c<{\egroup\egroup$}}

\def\gev{\;\hbox{GeV}}

\def\det{\hbox{det}}

\def\diag{\hbox{\diag}}
\def\zBB{{\mathbb{Z}}}
\def\lamh{\lambda_H}
\def\lams{\lambda_S}
\def\lamsm{\lambda_{SM}}
\def\z2{\zBB_2}
\def\mt{m_t}
\def\qs{Q^{\star}}
\def\mpl{M_{Pl}}
\def\mone{M_{h_1}}
\def\mtwo{M_{h_2}}
\def\zp{Z^\prime}
\def\mzp{M_{\zp}}
\def\vx{v_x}
\def\gx{g_x}
\def\xdd{\sigma_{\zp N}}
\def\msm{125.7}
\def\msmhalf{62.9}

\def\odm{\Omega_{DM} h^2}
\def\uone{U(1)_X}
\def\lsm{0.13}

\def\lsim{\mathrel{\raise.3ex\hbox{$<$\kern-.75em\lower1ex\hbox{$\sim$}}}}
\def\gsim{\mathrel{\raise.3ex\hbox{$>$\kern-.75em\lower1ex\hbox{$\sim$}}}}

\def\ot#1{%
  \mathrel{\vbox{\offinterlineskip\ialign{%
    \hfil##\hfil\cr
    $\scriptscriptstyle(\,\sim\,)$\cr
    \noalign{\kern-.1ex}
    $#1$\cr
}}}}

\def\inv#1{\frac1{#1}}
\begin{document}

\title{A stable Higgs portal with vector dark matter}

\author{M. Duch,}
\author{B. Grzadkowski}
\author{and M. McGarrie}
\affiliation{Faculty of Physics, University of Warsaw, Pasteura 5, 02-093 Warsaw, Poland}
\emailAdd{mateusz.duch@fuw.edu.pl}
\emailAdd{bohdan.grzadkowski@fuw.edu.pl}
\emailAdd{Moritz.McGarrie@fuw.edu.pl}
\date{\today}

\abstract{
We explore an extension of the Standard Model by an additional $U(1)$ gauge group and a complex scalar Higgs portal. As the scalar is charged under this gauge factor
this simple model supplies a vector dark matter candidate satisfying the observed relic abundance and limits from direct dark matter searches. An additional Higgs-like state, that may be heavier or 
lighter than the observed Higgs, is present and satisfies LEP and LHC bounds whilst allowing for absolute 
stability of the electroweak vacuum in a range of the parameter space.
}
\maketitle


\section{Introduction}
\label{sec:intro}
The discovery of a $125$~GeV scalar \cite{Aad:2012tfa,Chatrchyan:2012ufa} completes the Standard Model (SM) of particle physics and appears 
to confirm some rather intriguing features of nature. Firstly, the Higgs potential of the SM develops an instability at high field values, 
\cite{Cabibbo:1979ay,Hung:1979dn,Lindner:1985uk,Lindner:1988ww,Arnold:1989cb,Sher:1988mj,Sher:1993mf,Schrempp:1996fb}. Secondly, 
the theory on its own does not account for nearly 95\%  of the matter in the universe, 
known as dark matter (DM) \cite{Bertone:2004pz} and dark energy.

In this paper we explore a simple extension of the SM by an extra  $\uone$ gauge symmetry factor, supplemented  with 
an additional complex scalar charged solely under this $\uone$. It was studied by \cite{Lebedev:2011iq,Farzan:2012hh,Baek:2012se,Baek:2014jga} 
and by \cite{Hambye:2008bq,Gross:2015cwa,DiChiara:2015bua} in its non-Abelian version (for other realizations  of vector fields in the context of dark matter see \cite{Yu:2014pra,Ghorbani:2015baa}). The model can (if the real component of the complex scalar obtains a vev) generate a viable massive vector dark matter candidate. It also provides an additional Higgs mass eigenstate and introduces additional freedom in the theory to completely alleviate the issue of vacuum stability.

In this paper we use a collection of HEP-tools and software to give an accurate account of the currently viable parameter space of such a model. 
In particular we combine  \SARAH  \cite{Staub:2009bi,Staub:2010jh,Staub:2012pb,Staub:2013tta}, \PYRATE \cite{Lyonnet:2013dna} and \SPheno 
\cite{Porod:2011nf,Porod:2003um}  to explore stability at 2-loops and \MICRO \cite{Belanger:2001fz,Belanger:2004yn,Belanger:2006is,Belanger:2013oya} 
for a detailed study of vector dark matter for this model.  
We also compare our results to various experiments including LUX~\cite{Akerib:2013tjd} and XENON100~\cite{Aprile:2012nq}.

The paper is organized as follows. In section~\ref{sec:vdm} the model is defined and a detailed 
exposition of the parameters of the model is presented.
In section~\ref{rge} the renormalisation group equations of the theory are used to constrain the parameter space from physical consideration 
such as positivity, stability and the absence of Landau poles in the theory. 
In section~\ref{par_space} collider constraints are applied and the dark matter abundance constraint is 
imposed on the parameter space that remained. A detailed study of the DM-nucleon cross section is also performed.
Finally in section~\ref{summary} the findings are summarized.

In Appendix \ref{app:rel_abun}  we give a detailed account of the tree-level calculation of the  relic abundance.
Appendix \ref{app:modelfiles} supplies the \SARAH model files which fixes the conventions and allows for the generation of code for RGES 
and for the code used in \MICRO \cite{Belanger:2001fz,Belanger:2004yn,Belanger:2006is,Belanger:2013oya}.

\section{A model of vector boson dark matter}
\label{sec:vdm}

The vector dark matter (VDM)  model \cite{Hambye:2008bq,Lebedev:2011iq,Farzan:2012hh,Baek:2012se,Baek:2014jga} 
is an extension of the SM by an additional $\uone$ gauge symmetry together with a complex scalar field 
$S$,  whose vev generates a mass for this $U(1)$'s vector field. The quantum numbers of the scalar field are 
\beq
S  =   (0,{\bf 1},{\bf 1},1) \ \  \text{under}  \ \  U(1)_Y\times SU(2)_L \times SU(3)_c \times \uone.
\eeq
None of the SM fields are charged under the extra gauge group.  In order to ensure stability of the new vector boson
a $\z2$ symmetry is assumed to forbid $U(1)$-kinetic mixing between $\uone$ and  $U(1)_Y$. The extra gauge boson $A_\mu$  and 
the scalar $S$ field transform under $\z2$ as follows
\beq
A^{\mu}_X \rightarrow -A^{\mu}_X \ , \ S\rightarrow S^*,  \ \text{where} \  S=\phi e^{i\sigma}, \ \ {\rm so} \ \ \phi\rightarrow \phi,  \ \  \sigma\rightarrow -\sigma.
\eeq
All other fields are neutral under the $\z2$. 

The charge neutral vector bosons of $U(1)_Y$ \& $SU(2)_L$ $(B,W_3)$,  mix and the model leads to $(\gamma,Z)$.  The mixing matrix is taken to be
\begin{align} 
\left(\begin{array}{c} 
B\\ 
W_{{3 }}\end{array} \right) 
 = &  \left( 
\begin{array}{cc} 
\cos\theta_W  & -  \sin\theta_W      \\ 
 \sin\theta_W  & \cos\theta_W     
 \end{array} 
\right) 
\left(\begin{array}{c} 
\gamma\\ 
Z\end{array} \right)  
\end{align} 
where\footnote{$g'=\sqrt{3/5}g_1$  and $g=g_2$.} 
\beq
\cos\theta_W  =\frac{g}{\sqrt{g'^2+g^2}}  \ \  ,  \ \ 
\sin\theta_W  =\frac{g'}{\sqrt{g'^2+g^2}}.
\eeq 
At leading order the vector bosons masses are given by:
\beq
M_W=\inv2 g v, \ \ \ \ M_Z = \frac{1}{2}\sqrt{g^2+g'^2} v \ \ \ \text{and} \ \ \   \mzp = \gx \vx,
\eeq
where $v$ and $\vx$ are $H$ and $S$ vacuum expectation values (vev's): $(\langle H \rangle,\langle S \rangle)=\frac{1}{\sqrt{2}}(v,\vx)$.
The scalar potential for this theory is given by
\beq
V= -\mu^2_H|H|^2 +\lambda_H |H|^4 -\mu^2_S|S|^2 +\lambda_S |S|^4 +\kappa |S|^2|H|^2 .
\eeq
It will also be useful to define, for future reference, the parameter $\lambda_{SM}\equiv M_h^2/(2 v^2)=0.13$, 
where $M_h=\msm \gev$. 

The requirement of positivity for the potential implies the following constraints that we impose in all further discussions:
\beq
\lambda_H > 0, \ \ \lambda_S >0, \ \ \kappa > -2 \sqrt{\lambda_H \lambda_S}.
\label{positivity}
\eeq
Hereafter the above conditions will be referred to as the positivity or stability conditions. 

It is easy to find the minimization conditions for scalar fields 
(without losing generality one can assume $v,\vx>0$):
\beq
(2\lambda_H v^2 + \kappa \vx^2 - 2\mu^2_H) v = 0\ \  \text{and} \ \  (\kappa v^2 + 2\lambda_S v^2_x - 2\mu^2_S)\vx = 0
\label{min_con}
\eeq
If $\mu_{H,S}^2<0$ the global minimum at $(0,0)$ is the only extremum. For $\mu_{H,S}^2>0$ the point $(0,0)$ is a local maximum of the potential,
in this case $(0,\frac{\mu_S}{\sqrt{\lambda_S}})$ and $(\frac{\mu_H}{\sqrt{\lambda_H}},0)$ are global minima if 
$\kappa^2>4\lambda_H\lambda_S$, otherwise they are saddle points and the global minima are determined by
\beq
v^2=\frac{4 \lambda_S \mu^2_H - 2\kappa \mu^2_S }{4\lambda_H\lambda_S-\kappa^2},\ \ \vx^2=\frac{4 \lambda_H \mu^2_S - 2\kappa \mu^2_H }{4\lambda_H\lambda_S-\kappa^2}.
\eeq
For the VDM model only the latter case is relevant, since both vevs need to be non-zero to give rise to the masses of the Standard Model fields and dark vector boson. Both scalar fields can be expanded around corresponding vev's as follows
\beq
S=\frac{1}{\sqrt{2}}(\vx+ \phi_S +i\sigma_S)  \ \ , \ \  H^0= \frac{1}{\sqrt{2}}(v + \phi_H+ i\sigma_H)   \ \  \text{where} \ \  H=\binom{H^+}{H^0}.
\eeq
The mass squared matrix $\mathcal{M}^2$ for the fluctuations $ \left(\phi_H, \phi_S\right)$ and their eigenvalues read
\begin{equation} 
\mathcal{M}^2 = \left( 
\begin{array}{cc}
2 \lambda_H v^2  & \kappa v \vx \\ 
 \kappa v \vx &2 \lambda_S v^2_x 
\end{array} 
\right) \ \ , \ \ 
M^2_\pm=\lambda_H v^2 +\lambda_S \vx^2\pm \sqrt{\lambda_S^2 \vx^4 - 2\lambda_H\lambda_S v^2 \vx^2 + \lambda_H^2 v^4 +\kappa^2 v^2 \vx^4}.
\label{massmatrix}
 \end{equation} 
The matrix $\mathcal{M}^2$ could be diagonalized by the orthogonal rotation $R$, such that 
$\mathcal{M}_{\text{diag}}^2= R^{-1} \mathcal{M}^2 R$. The convention adopted for the ordering of the
eigenvalues and for mixing angle $\alpha$ is the following  
\begin{equation} 
\mathcal{M}_{\text{diag}}^2 = \left( 
\begin{array}{cc}
\mone^2  &0 \\ 
0 &\mtwo^2
\end{array} 
\right)   ,
 \ \   \ 
R = \left( 
\begin{array}{cc}
\cos \alpha   & -\sin \alpha \\ 
\sin \alpha &\cos \alpha
\end{array} 
\right), 
 \ \   \
 \left( 
\begin{array}{c}
h_1\\ 
h_2
\end{array} 
\right) = R^{-1}
\left(
\begin{array}{c}
\phi_H\\ 
\phi_S
\end{array} 
\right),
\end{equation} 
where $\mone=\msm\gev$ is the mass of the observed Higgs particle. Then we obtain
\beq
\sin 2\alpha =\frac{\text{sign}(\lamsm - \lamh)\,2\mathcal{M}_{12}^2}{  \sqrt{ (\mathcal{M}^2_{11}-\mathcal{M}^2_{22} )^2 +4(\mathcal{M}^2_{12})^2}    }, \ \ \ 
\cos 2\alpha = \frac{\text{sign}(\lamsm-\lamh)(\mathcal{M}_{11}^2 -\mathcal{M}_{22}^2) }{  \sqrt{ (\mathcal{M}^2_{11}-\mathcal{M}^2_{22} )^2         +4(\mathcal{M}^2_{12})^2}    }.
 \label{anglesformulas}
\eeq
Note that since vev of $H$, if fixed at $246.22\gev$, with $\kappa=0$ (no mass mixing) and $\lamh \neq \lamsm$ it is only $h_2$ which can have the observed Higgs mass of $\msm\gev$. 
Even though the mass matrix is diagonal in this case, however in order to satisfy our convention that $\mone=\msm\gev$
a rotation by $\alpha=\pm\pi/2$ is required in such a case. 

There are 5 real parameters in the potential: $\mu_H$, $\mu_S$, $\lamh$, $\lams$ and $\kappa$. Adopting the minimization conditions
(\ref{min_con}) $\mu_H$, $\mu_S$ could be replaced by $v$ and $\vx$. The SM vev will be fixed at $v=246.22\gev$. 
Using the condition $M_{h_1} = \msm\gev$, $\vx^2$ could be eliminated via (\ref{massmatrix})
in terms of $v^2,\lamh,\kappa,\lams,\lamsm = M_{h_1}^2/(2 v^2)$:
\beq
\vx^2=v^2\frac{4\lamsm (\lamh-\lamsm)}{4\lams (\lamh-\lamsm)-\kappa^2} 
\label{vx}
\eeq
Therefore eventually there are 4~independent unknown parameters in the model 
$(\lambda_H, \kappa,\lambda_S, \gx)$, where $\gx$ is the $\uone$ coupling constant.

It is important to notice that positivity of $\vx^2$ implies for $\lamh>\lamsm$ that
\beq
\lamh > \frac{\kappa^2}{4\lams}+\lamsm
\label{lamh_lim}
\eeq
Applying the fact that $M_{h_1}^2 +  \mtwo^2 = 2(\lamh v^2+\lams \vx^2)$ together with (\ref{vx}) one finds the
following universal formula for the mass of the non-standard Higgs:
\beq
\mtwo^2=v^2\frac{2(\lamh-\lamsm)(4\lamh\lams-\kappa^2)}{4\lams(\lamh-\lamsm)-\kappa^2}
\label{non_sm_mass}
\eeq 
It is easy to see that positivity of $\mtwo^2$ is guaranteed if the following conditions are satisfied:
\bit
\item for $\lamh < \lamsm$, $\lamh > \frac{\kappa^2}{4\lams}$,
\item for $\lamh > \lamsm$, $\lamh > \frac{\kappa^2}{4\lams}+\lamsm$ (same as (\ref{lamh_lim})).
\eit

\begin{figure}[h!]\centering
\includegraphics[width=.8\textwidth]{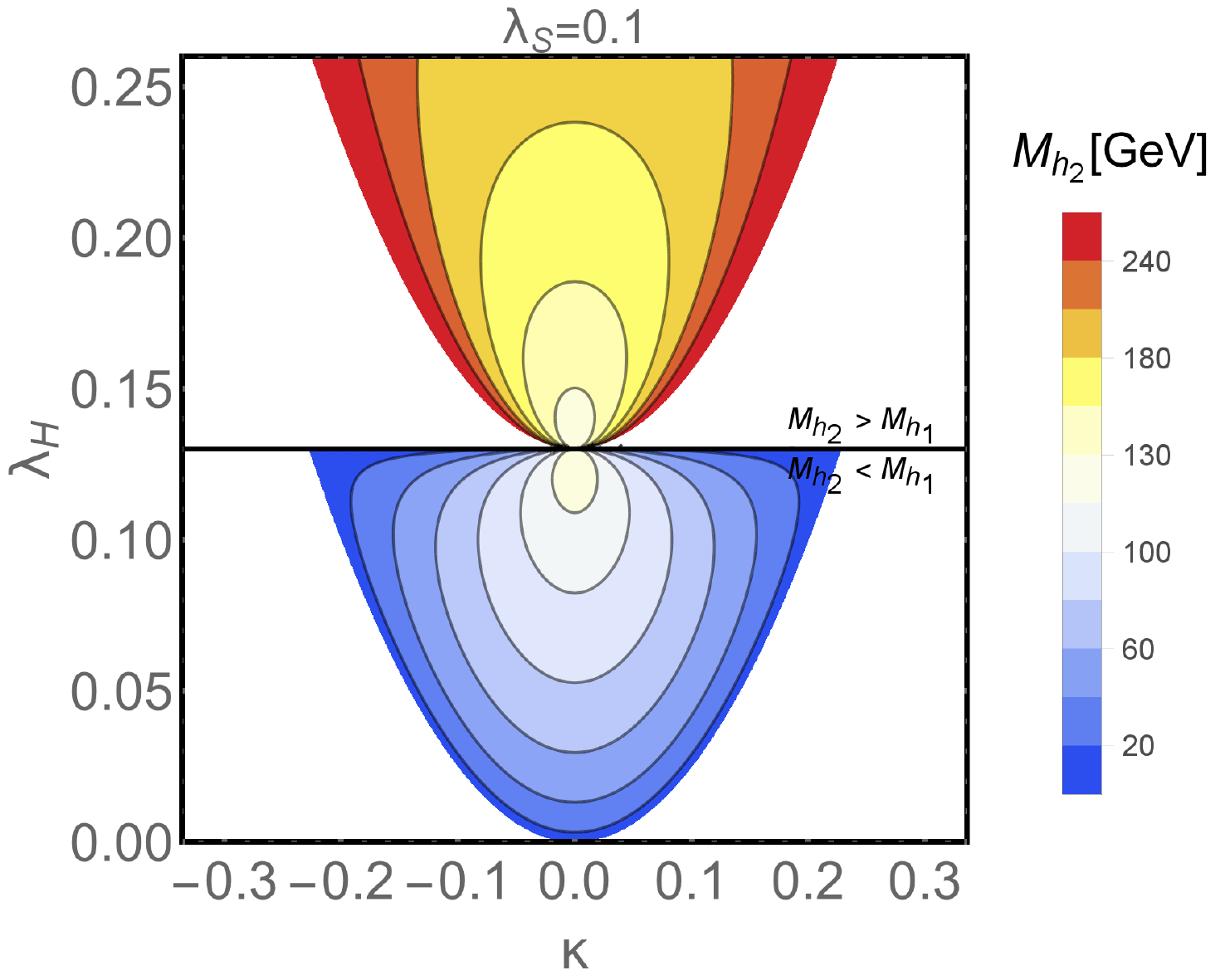}
\caption{Contour plots for masses of the non-standard ($h_2$) Higgs particle in the plane $(\lambda_H,\kappa)$ 
for fixed value $\lambda_s=0.1$. 
The observed Higgs mass is set to be $M_{h_1}=\msm\gev$, the vev of the Higgs doublet is fixed at $v=246.22\gev$. In the bottom part of the plot
($\lambda_H<\lamsm = M_{h_1}^2/(2 v^2)=\lsm$) the heavier Higgs is the currently observed one,  while in the upper part ($\lambda_H>\lambda_{SM}$) 
the lighter state is the observed one. The horizontal black line $\lambda_H=\lambda_{SM}$ separates the two scenarios. White regions in the
upper and lower parts are disallowed by the positivity conditions for $\vx^2$ and $\mtwo^2$, respectively. }
\label{mass_plot}
\end{figure}

From (\ref{massmatrix}) one can easily derive the following useful inequalities
\beq
M^2_+\geq\lambda_H v^2 +\lambda_S \vx^2 + |\lambda_H v^2 - \lambda_S \vx^2| = \text{Max}(2\lambda_H v^2,2\lambda_S \vx^2),
\eeq
\beq
M^2_-\leq\lambda_H v^2 +\lambda_S \vx^2 - |\lambda_H v^2 - \lambda_S \vx^2| = \text{Min}(2\lambda_H v^2,2\lambda_S \vx^2).
\eeq
Therefore for $\lambda_H>\lamsm$ we have $M_+>M_{h_1}$, so the lighter scalar is the SM-like Higgs particle, while 
when $\lambda_H<\lambda_{SM}$ the observed Higgs particle is the heavier one. Consequently $\lambda_H$ is the sole 
parameter that distinguishes between these two scenarios.
To illustrate that behavior we show in fig.~\ref{mass_plot} contours of non-standard Higgs masses in the plane $(\kappa, \lamh)$.
Note the presence of $\mtwo^2=0$ contour (the most external blue contour in the lower part of the figure) that corresponds 
to vanishing determinant of the matrix (\ref{massmatrix}): $(\kappa^2-4\lamh\lams) \vx^2=0$. In fact it is interesting to consider the special case
of $\kappa^2-4\lamh\lams=0$, then if one requires that the potential has a minima for $v,\vx\neq 0$ it is necessary to assume 
that $\mu_H^2/\lamh^{1/2}=\mu_S^2/\lams^{1/2}$, which implies that the potential could be written as
\beq
-\tilde\mu^2\left(\lamh^{1/2}|H|^2 + \lams^{1/2}|S|^2\right) + \left(\lamh^{1/2}|H|^2 + \lams^{1/2}|S|^2\right)^2
\label{spec_pot}
\eeq
with $\tilde\mu^2\equiv\mu_H^2/\lamh^{1/2}$. This potential has equipotential contours (in the unitary gauge) on ellipses such that 
$\lamh^{1/2}|H|^2 + \lams^{1/2}|S|^2=$ constant. Fluctuation along the ellipses that corresponds to the minimum is the massless mode.
Note that this is a mode that exists even if $\vx\neq 0$, so this parameter choice is different from the case discussed below 
at the end of this section where $\vx$ is approaching $0$ (so $\det (\mathcal{M}^2) \to 0$ as well).

\begin{figure}[h!]\centering
\includegraphics[width=.45\textwidth]{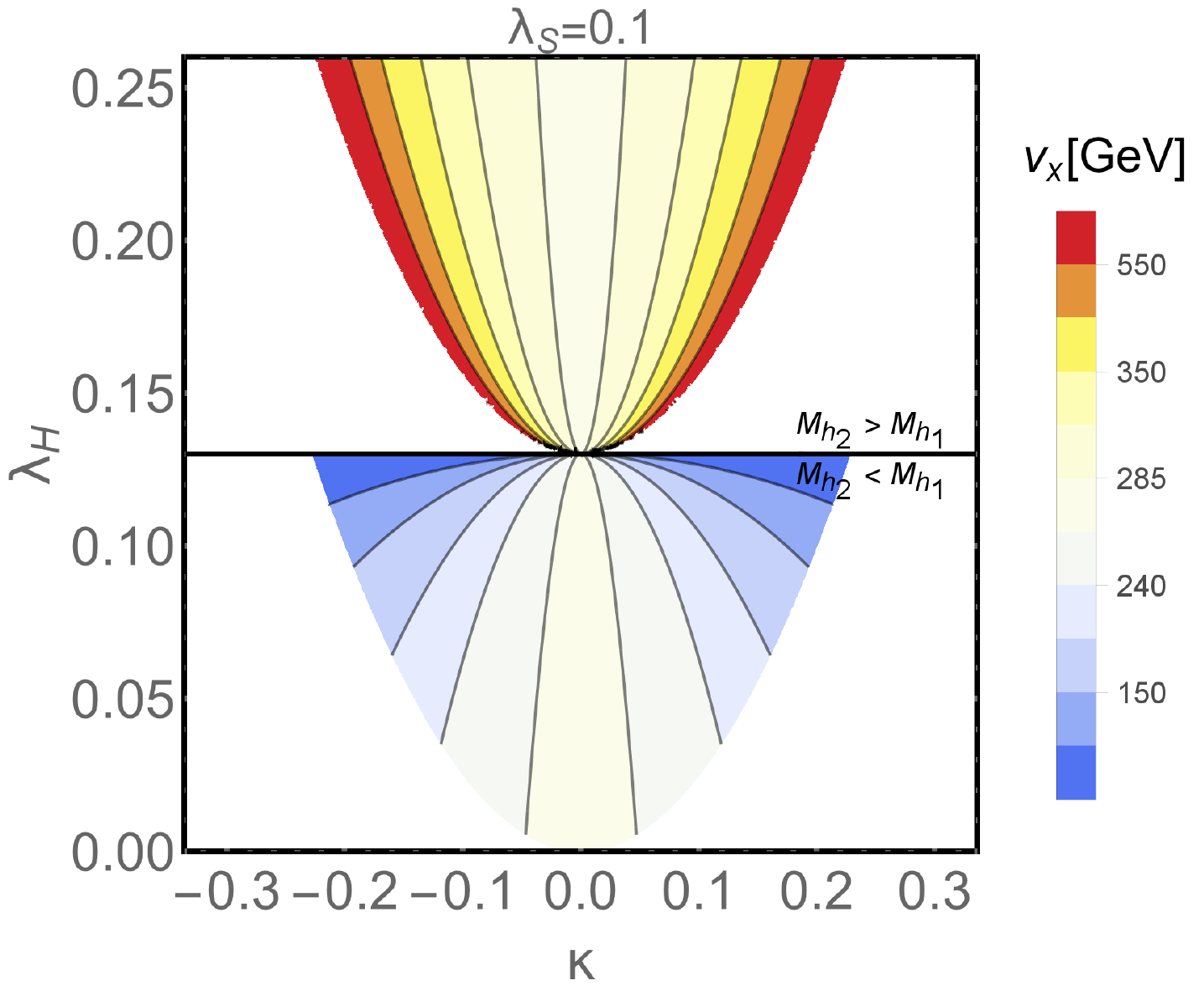}
\includegraphics[width=.45\textwidth]{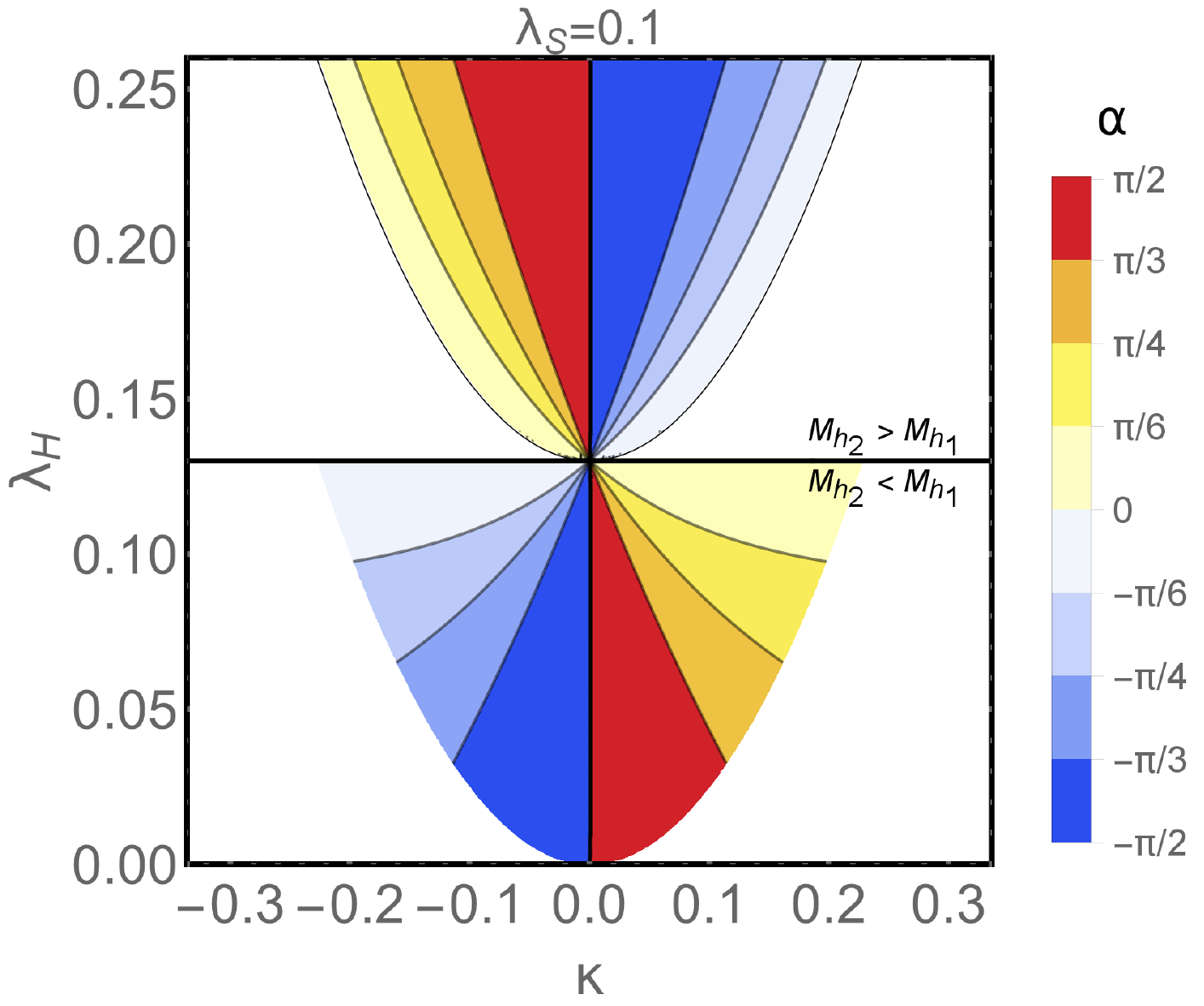}
\caption{Contour plots for the vacuum expectation value of the extra scalar $\vx\equiv \sqrt{2}\langle S \rangle$ (left panel)
and of the mixing angle $\alpha$ (right panel) in the plane $(\lambda_H,\kappa)$ for fixed value $\lambda_s=0.1$. 
In the bottom part of the plots the positivity of $\mtwo^2$ limits the allowed area
by $\lamh > \kappa^2/(4\lams)$, so the white region is disallowed.   
The observed Higgs mass is set to be $M_{h_1}=\msm\gev$, the vev of the Higgs doublet is fixed at $v=246.22\gev$. 
In the bottom part of the plots ($\lambda_H<\lambda_{SM}$) the heavier Higgs is the currently observed one,  
while in the upper part ($\lambda_H>\lambda_{SM}$) the lighter state is the observed one. 
The horizontal black line $\lambda_H=\lambda_{SM}=\lsm$ separates the two scenarios.  }
\label{vx_plot}
\end{figure}

The behavior of $\vx$ is presented in fig \ref{vx_plot}. For $\lambda_H<\lambda_{SM}$ and $\lamh > \frac{\kappa^2}{4\lams}$ (as required by
the scalar mass positivity) one finds
\beq
 0 < \vx^2 < v^2 \frac{\lamsm}{\lams}
\label{vx_lim_down}
\eeq
Similarly for $\lambda_H>\lambda_{SM}$ and $\lamh > \kappa^2/(4\lams)+\lamsm$ 
$\vx$ is limited by
\beq
v^2 \frac{\lamsm}{\lams} < \vx^2 < \infty
\label{vx_lim_up}
\eeq
and it is diverging at the parabola $\lambda_H = \kappa^2/(4\lams) + \lamsm$. For $\lamh<\lamsm$ only the physical region
corresponding to $\mtwo^2>0$ ($4\lamh\lamsm - \kappa^2>0$) is shown in fig \ref{vx_plot}.

In order to understand the behaviour of the mixing angle $\alpha$ one can directly adopt the formulae (\ref{anglesformulas}). 
Since $\alpha$ varies in the range $[-\pi/2,\pi/2]$, the absolute value of $|\alpha|$ and its sign can be read from 
the inverse of $\cos(2\alpha)$ and $\sin(2\alpha)$, respectively. 
The coupling of $h_1$ eigenstate (the one that has $\msm\gev$ mass) to $VV$ is proportional to $\cos\alpha$ 
therefore the LHC data favours regions of $\alpha \sim 0$. As can be found from 
(\ref{anglesformulas}) small $\alpha$ corresponds to either to $\vx\to \infty$ (for $\lamh>\lamsm$) or
to $\vx \to 0$ (for $\lamh<\lamsm$). One can see the same behaviour from fig.~\ref{vx_plot}.

It is worth investigating the SM limit of the VDM model. Fig.~\ref{vx_plot} is a good starting point
as it is easy to recognize regions in the parameter space that imply vanishing corrections (relative to the SM)
to $h_1$ couplings (this is what we define by the SM limit). So, for $\lamh > \lamsm$ this is the parabola 
$\lamh = \frac{\kappa^2}{4\lams}+\lamsm$ where $\vx\to \infty$ and $\alpha$ vanishes, while for $\lamh < \lamsm$
it is the vicinity of $\lamh=\lamsm$. The case $\lamh > \lamsm$  is less interesting as it is just the decoupling
limit with $\mtwo \sim 2\lams \vx^2 +\cdots \to \infty$, see fig.~\ref{mass_plot}, 
when the effective low energy theory is just the SM with $h_2$ being integrated out. 
The mixing angle $\alpha$ behaves in that region as $\alpha \sim \kappa/\lams (v/\vx)^2 + \cdots$.
More interesting is the region with $\lamh$ approaching $\lamsm$ from below, 
as there the mass of $h_2$ goes to zero (also $\vx \to 0$), as seen in fig.~\ref{vx_plot}.
In this region the model contains a SM-like scalar with the mass of $\msm$ and almost massless
state ($h_2$) that decouples from $VV$, $h_1h_1$ and fermions, however its cubic ($\kappa h_1 h_2^2$) and quartic
($\kappa h_1^2 h_2^2$) scalar couplings remain. This limit of the VDM model turns out to be phenomenologically
unattractive since the Higgs boson $h_1$ would decay invisibly into pairs $h_2h_2$.

\section{The renormalisation group equations}
\label{rge}

As will be illustrated shortly, in order to investigate vacuum stability in the VDM model it is necessary to use 2-loop RGEs. 
We have adopted \SARAH \cite{Staub:2009bi,Staub:2010jh,Staub:2012pb,Staub:2013tta} to obtain the full  set of 2-loop beta functions.
However, in order to facilitate further discussion we show below the beta functions of the gauge couplings at 1-loop level, even though
whenever we impose any constraints on the parameter space that rely on the RGE running we always use 2-loop beta functions. 
\beq
 \beta_{g_a}=16\pi^2\frac{d}{dt}g_a
\eeq
we have
\beq
\beta_{g_1}^{(1)}  =  
\frac{41}{10}  g_{1}^{3}  \ \ , \ \ 
\beta_{g_{x}}^{(1)}  =  
\frac{1}{3}  g_{x}^{3} \ \  , \  \
\beta_{g_2}^{(1)}  =  
-\frac{19}{6} g_{2}^{3} \phantom{sp}\text{and} \phantom{sp}
\beta_{g_3}^{(1)}  =  
-7 g_{3}^{3} 
\eeq
The various Higgs quartic couplings at 1-loop run as 
{\allowdisplaybreaks  \begin{align} 
\beta_{\lambda_H}^{(1)} & =  
\frac{27}{200} g_{1}^{4}  +\frac{9}{20} g_{1}^{2} g_{2}^{2} +\frac{9}{8} g_{2}^{4}  -\frac{9}{5} g_{1}^{2} \lambda_H -9 g_{2}^{2} \lambda_H +24 \lambda_{H}^{2}+\kappa^{2}   \nonumber \\ 
 &\phantom{00}  
-6 \mbox{Tr}\Big({Y_u  Y_{u}^{\dagger}  Y_u  Y_{u}^{\dagger}}\Big)  -2 \mbox{Tr}\Big({Y_e  Y_{e}^{\dagger}  Y_e  Y_{e}^{\dagger}}\Big)  -6 \mbox{Tr}\Big({Y_d  Y_{d}^{\dagger}  Y_d  Y_{d}^{\dagger}}\Big) 
 \nonumber \\ 
 &\phantom{0000000}  
+12 \lambda_H \mbox{Tr}\Big({Y_d  Y_{d}^{\dagger}}\Big) +4 \lambda_H \mbox{Tr}\Big({Y_e  Y_{e}^{\dagger}}\Big) +12 \lambda_H \mbox{Tr}\Big({Y_u  Y_{u}^{\dagger}}\Big)    \label{beta_lamh} \\  
\beta_{\lambda_S}^{(1)} & =  
\frac{1}{2} \Big(-36 g_{x}^{2}\lambda_S  + 27 g_{x}^{4}   + 40 \lambda_{S}^{2}  + 4 \kappa^{2} \Big) \label{beta_lams} \\
\beta_{\kappa}^{(1)} & =  
\frac{\kappa}{10} \Big[ -9 g_{1}^{2}  -90 g_{x}^{2}   -45 g_{2}^{2} +120 \lambda_H +80 \lambda_S  +40 \kappa   \nonumber \\ 
 &\phantom{spacespacespace}+60 \mbox{Tr}\Big({Y_d  Y_{d}^{\dagger}}\Big) +20  \mbox{Tr}\Big({Y_e  Y_{e}^{\dagger}}\Big) +60 \mbox{Tr}\Big({Y_u  Y_{u}^{\dagger}}\Big) \Big]\label{beta_kappa} 
\end{align}} 
Above $Y_{u,d,e}$ denote the corresponding Yukawa matrices. Since here we are mainly concerned with the case of the masses for the extra scalar $h_2$ and $\zp$ of the order of
the electroweak scale and since initial conditions for RGE running will be specified at $Q=\mt$ (relatively large scale) therefore we will adopt the above beta functions 
neglecting decoupling of extra degrees of freedom below the scale $Q=\mtwo, \mzp$.

We have verified that, in the SM limit, our 2-loop running of $\lamh(Q)$
agrees with known results~\cite{Buttazzo:2013uya}.

In order to explore the stability and positivity of the theory we used the two-loop RGEs and the tree-level potential. 
To improve the precision of this work further
one would likely need the one-loop improved effective potential before extending to 3-loop RGEs.

\begin{figure}[h]
\begin{center}
\includegraphics[width=.45\textwidth]{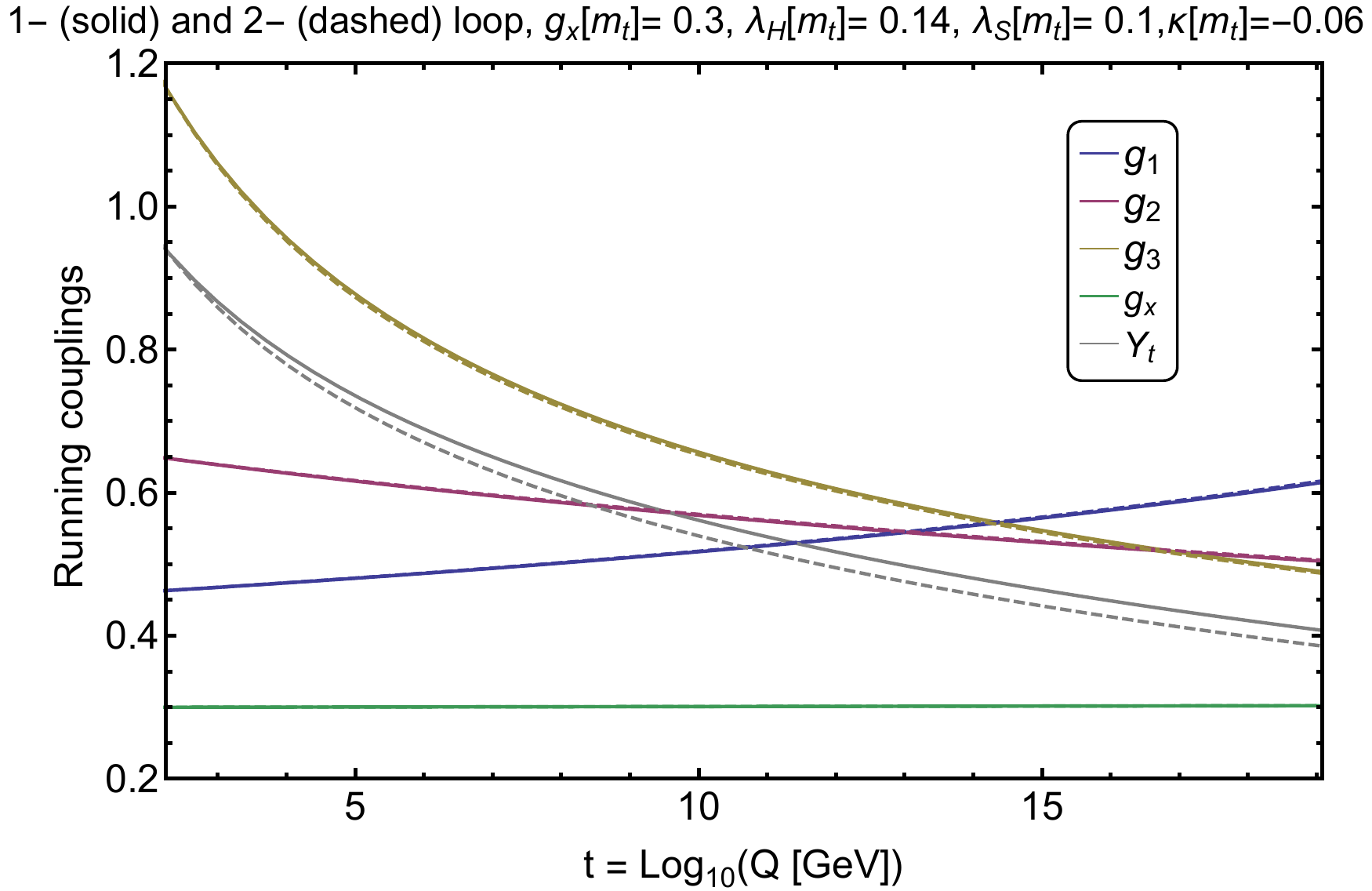}\qquad
\includegraphics[width=.45\textwidth]{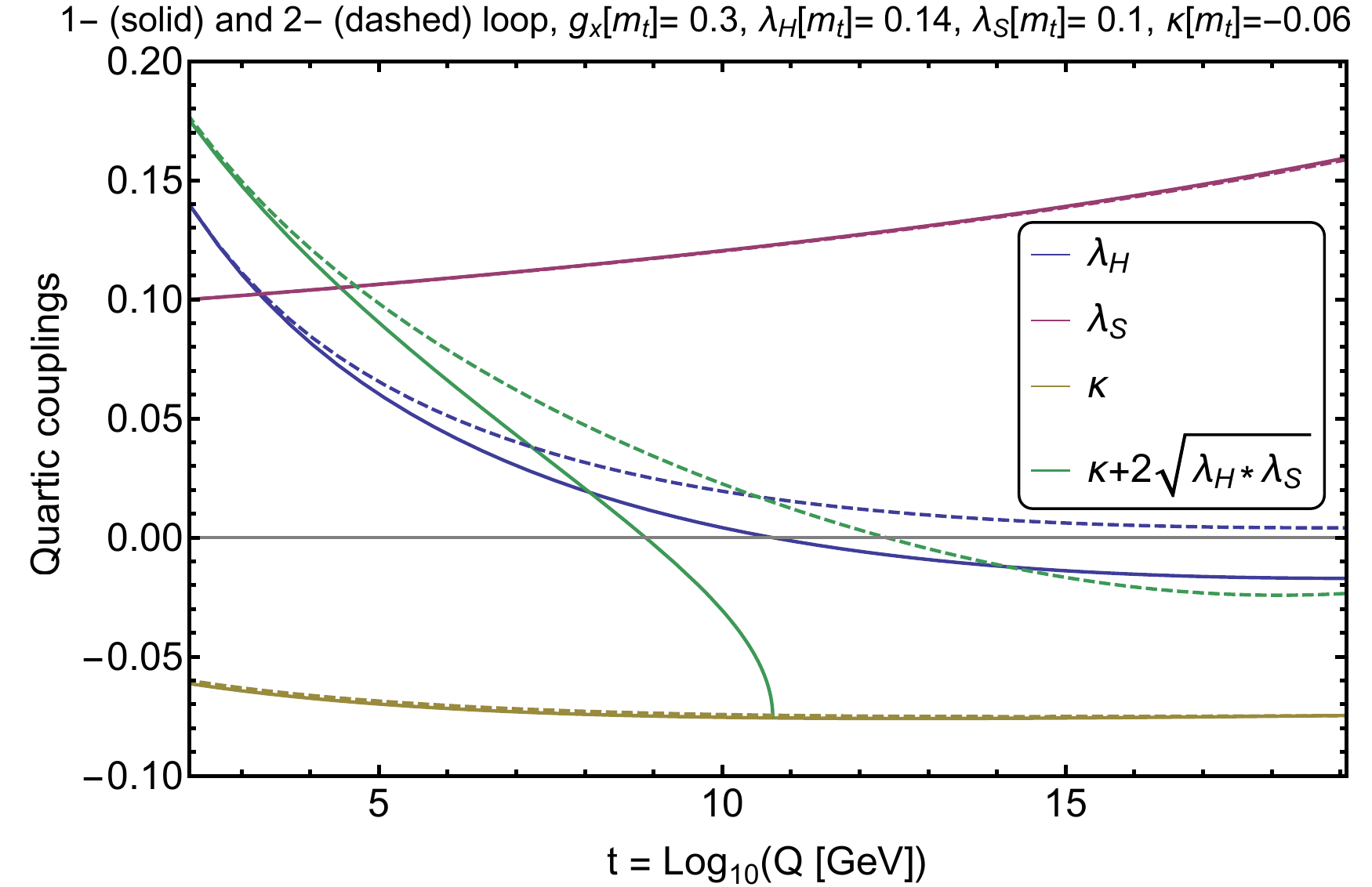}\qquad
\caption{Running of various parameters at 1- and 2-loop, in solid and dashed lines respectively. For this choice of parameters 
$\lambda_H(Q)>0$ at 2-loop (right panel blue) but not at 1-loop. $\lams(Q)$ is always positive (right panel red), running of $\kappa(Q)$ is very limited, 
however the third positivity condition $\kappa(Q) + 2\sqrt{\lamh(Q)\lams(Q)}>0$
is violated at higher scales even at 2-loops (right panel green).  }
\end{center}
\label{fig:rges}
\end{figure}

An example of a representative point in the parameter space at $Q=\mt$ is pictured in figure \ref{fig:rges}, 
where we show both 1- and 2-loop running of the gauge and scalar quartic couplings.  A few comments are here in order.
The running of gauge couplings (left panel of \ref{fig:rges}) is rather stable and similar for 1- and 2-loop beta functions.
However the Yukawa coupling $Y_t$ already shows (left panel) some sensitivity to the approximation adopted for the beta functions.
It is important to note the relevance of 2-loop running of $\lamh(Q)$; as seen in the right panel of fig.~\ref{fig:rges} the scale of
instability (i.e. a scale $\qs$ from which on  $\lamh(Q)$ is negative) is very sensitive to the RGE running precision.
For 1-loop beta functions $\qs\sim 10^{11}\gev$ while for 2-loop approximation $\lamh(Q)$ remain positive up to scale $10^{19}\gev$.
Other quartic couplings ($\kappa(Q)$ and $\lams(Q)$) do not require 2-loop beta functions as their evolution  is nearly the same for 
1- and 2-loop beta functions. It is worth noticing that $\beta_{\lams}$ is always positive, so the stability condition $\lams>0$ 
can never be violated radiatively.  The evolution of $\kappa$ is rather mild as $\beta_\kappa \propto \kappa$ therefore at least for small
$\kappa$ the evolution is quite suppressed. 
In the right panel of fig.~\ref{fig:rges} we also show the running of the positivity condition
$\kappa(Q) + 2\sqrt{\lamh(Q)\lams(Q)}>0$. 
Here the 2-loop effects are again important. The 1-loop curve terminates already
around $Q\sim 10^{11}\gev$, the scale at which 1-loop $\lamh(Q)$ becomes negative, so that the positivity condition $\kappa(Q) + 2\sqrt{\lamh(Q)\lams(Q)}>0$
can not be verified. On the other hand the 2-loop running of this condition shows that it is satisfied up to $Q\sim 10^{12}\gev$.
The choice of initial conditions for this plot illustrates the fact that there exist initial conditions (i.e. points in the parameter 
space $(\lambda_H(\mt), \kappa(\mt),\lambda_S(\mt), \gx(\mt))$) such that even though the condition 
$\kappa(Q) + 2\sqrt{\lamh(Q)\lams(Q)}>0$ is satisfied at low scale it fails at high energies.

\subsection{Stability}
\label{sta_pos}

\begin{figure}[h]\centering
\includegraphics[width=1\textwidth]{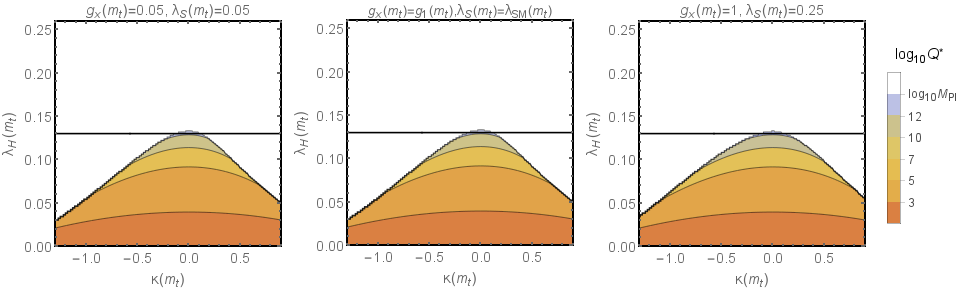}
\caption{The stability frontier for the $H$ direction: these plots identify the renormalisation scale $t^*=Log_{10}(Q^*)$ at which $\lambda_H(Q^*)=0$ and the vacuum 
becomes unstable, as a function of $(\lambda(\mt),\kappa(\mt))$.  The horizontal solid black line corresponds to $\lamh(\mt)=\lamsm\simeq\lsm$.
For $t=Log_{10}(Q^*)>Log_{10}(M_{Pl})=19.09$, the vacuum is absolutely 
stable up to that scale. At lower values the vacuum is meta-stable however a lower scale does not imply increased instability, 
one must further determine the tunneling rate. }
\label{lambdahplots}
\end{figure}

The constraints (\ref{positivity}) can be used to determine areas of parameter space \newline
$(\lambda_H(\mt), \kappa(\mt), \lambda_S(\mt), \gx(\mt))$ 
in which the conditions for stability/positivity of the potential are satisfied at all renormalization scales.
In the SM, the absolute stability is ensured just by the positivity of the quartic coupling at all energy scales $Q$: $\lamh(Q)>0$.
Since the mass of the Higgs boson is known experimentally the initial condition for running of $\lamh(Q)$ is fixed as
$\lamh(\mt)=M_{h_1}^2/(2 v^2)=\lamsm=\lsm$
and for this initial value
$\lamh(Q)$ becomes negative at some scale causing the instability.
However here, in the presence of the extra scalar $S$ this is not necessarily the case; the LHC Higgs mass measurement fixes 
the following combination of couplings and vev's:
\beq
M_{h_1}^2=\lambda_H v^2 +\lambda_S \vx^2\pm \sqrt{\lambda_S^2 \vx^4 - 2\lambda_H\lambda_S v^2 \vx^2 + \lambda_H^2 v^4 +\kappa^2 v^2 \vx^4}.
\label{mh_con}
\eeq
It is easy to see that the VDM model has the freedom to increase the value of $\lamh$ at low scales; a freedom which the SM does not possess.
Larger initial values of $\lamh$ such that $\lamh(\mt)>\lamsm$ are allowed delaying the instability (by shifting up the scale at which $\lamh(Q) < 0$).
There is also another remedy for the instability within the VDM model; even if the initial $\lamh$ is smaller than its SM value,
$\lamh(\mt)<\lamsm$, still there is a chance to lift the instability scale if appropriate initial value of the portal coupling
$\kappa(\mt)$ is chosen. This effect is caused by the positive $\kappa^2$ contribution to the beta function $\beta_{\lamh}$ that partially 
compensates the negative top-quark effect, see (\ref{beta_lamh}). Fig.~\ref{lambdahplots} illustrates the way the SM stability
problem encoded by $\lamh(Q)<0$ for $Q>\qs$ could be relaxed within the VDM model. The white region above the horizontal line $\lamh(\mt)=\lamsm$
shows the region of $\lamh(\mt)>\lamsm$ so the positivity of $\lamh$ up to the Planck scale could be easily guaranteed. On the other hand
the white region below the line $\lamh(\mt)=\lamsm$ shows those pairs of $(\lamh,\kappa)$ for which even though the starting point for
$\lamh$ evolution is lower than for the SM, nevertheless the extra positive contribution to $\beta_{\lamh}$ makes $\lamh$ positive 
up to the Planck scale. Clearly for large $\kappa$ the stability region increases (for negative $\kappa$ the other stability condition gives tighter constraint). The colorful regions show the scale at which 
$\lamh$ becomes negative. The three panels shown in fig.~\ref{lambdahplots} 
correspond to three different pairs of initial values for
$(\gx(\mt),\lams(\mt))$. As seen, the sensitivity to those choices is very weak even though $\gx(\mt)$ and $\lams(\mt)$ vary in a wide range,
in fact this is understandable since the evolution of $\lamh(Q)$ is influenced by $\gx(\mt)$ and $\lams(\mt)$ only indirectly through the 
presence of $\kappa^2$ in the beta function $\beta_{\lamh}$, see  (\ref{beta_lamh}).

\begin{figure}[h]\centering
\includegraphics[width=1\textwidth]{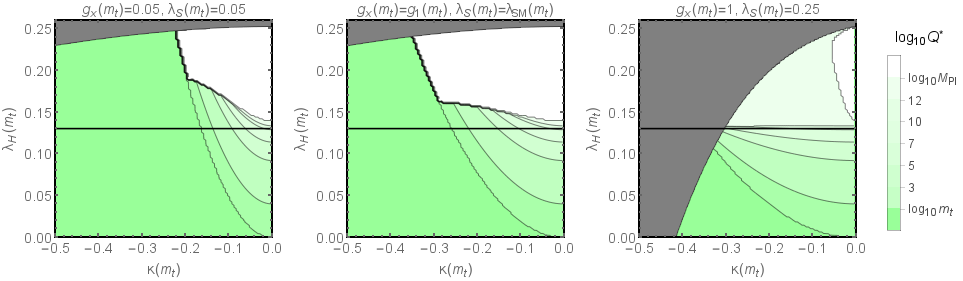}
\caption{ The ``in between'' stability  frontier :  these plots identify the scale $t^*=Log_{10}(Q^*)$ at which the positivity condition 
$\kappa(Q) + 2\sqrt{\lamh(Q) \lams(Q)}$ fails and the vacuum becomes unstable, as a function of $(\lambda(\mt),\kappa(\mt))$
for fixed choices of $(\gx(\mt),\lams(\mt))$ specified above each panel. The horizontal solid black line corresponds to $\lamh(\mt)=\lamsm\simeq\lsm$.
The gray area is excluded by the requirement that there is no Landau poles up to the Planck mass. }
\label{kappaplots}
\end{figure}

In the VDM model the SM stability problem (positivity of $\lamh$) is easily solved as was illustrated above. However in this case  
positivity requires two extra constraints: $\lams>0$ and $\kappa + 2\sqrt{\lamh \lams}>0$. Since $\beta_{\lams}>0$ therefore
whenever $\lams(\mt)>0$ the positivity is preserved during the evolution. However the second extra condition is non-trivial, 
as illustrated in fig.~\ref{fig:rges} it is possible that $\kappa(Q) + 2\sqrt{\lamh(Q) \lams(Q)}$ changes sign while running
from low energies up. Fig.~\ref{kappaplots} shows the scale at which $\kappa(Q) + 2\sqrt{\lamh(Q) \lams(Q)}$  becomes negative
as a function of $(\lambda(\mt),\kappa(\mt))$ for three fixed sets of $(\gx(\mt),\lams(\mt))$. 

Stability of the $\uone$ dark matter model was also discussed in \cite{DiChiara:2014wha}. The vacuum stability induced by the dark matter has also been 
considered in the context of complex~\cite{Gonderinger:2012rd} and real~\cite{Alanne:2014bra,Khan:2014kba} extra scalars serving as dark matter candidates.

\subsection{The Landau poles}
\label{lan_pol}

\begin{figure}[h]\centering
\includegraphics[width=1\textwidth]{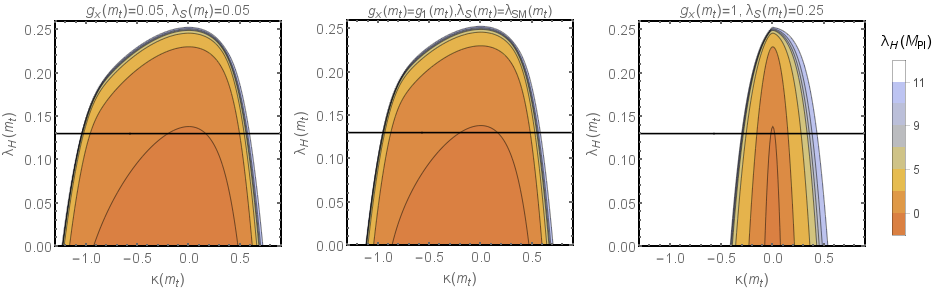}
\caption{Contour plots of $\lamh (M_{Pl})$ in the plane of $(\lambda(\mt),\kappa(\mt))$ for fixed $\gx(\mt)$ and $\lams(\mt)$
specified above each panel. The horizontal solid black line corresponds to $\lamh(\mt)=\lamsm\simeq\lsm$.
The plots allow one to identify regions (white) in which the $\lamh(Q)$ Landau pole is below the Planck scale.}
\label{landauplots}
\end{figure}

As we have discussed above, the additional freedom in the Higgs sector, that is due to the presence of the Higgs portal $\kappa |S|^2|H|^2$
allows one to increase the low scale value of $\lamh$ sufficiently to avoid its negative value (instability in the $H$ direction) at high scales. 
However, this possibility is bounded from above by the requirement that 
there are no Landau poles in the evolution of $\lamh$ (or any other parameter - a pole in the evolution of any coupling implies
divergence of all of them at the same energy) up to a chosen high scale, e.g. the Planck mass.  
In fig.~\ref{landauplots} we show contour plots of $\lamh(\mpl)$ in $(\lamh(\mt),\kappa(\mt))$ space for fixed $\gx(\mt)$ and $\lams(\mt)$.
It is clear that too large $\kappa(\mt)$ or $\lamh(\mt)$ implies early divergence of $\lamh(Q)$. Also when  $\gx(\mt)$ and/or $\lams(\mt)$
grow (from left to the right panel) the safe region shrinks in agreement with expectations. 

A few comments are in order here. Quartic scalar couplings $\lamh$, $\lams$, $\kappa$ and $\gx$ are free parameters in the model.
However for perturbative expansion to be valid their values can not be too large since otherwise 
the expansion has no chance to converge. The maximum adopted by various authors is to some extent
subjective and usually varies between $1$ and $4\pi$. Here we have the advantage of knowing both
1- and 2-loop beta functions for the RGE evolution of the couplings, therefore the relevance
of the 2-loop could be quantitatively estimated for different (large) values of couplings. 
We found that requiring the 2-loop correction to be smaller than $100\%$ of the 1-loop result for any quartic scalar coupling, 
i.e. $|(\lambda_i^{(2)} - \lambda_i^{(1)})/\lambda_i^{(1)}| < 1$, implies that the coupling should 
not exceed a value close to $2\pi$.  
Therefore, in the numerical results, whenever it is stated that
a Landau pole appears, it is meant that the corresponding coupling reaches a value of $2\pi$. 
For larger couplings the 2-loop contributions start to dominate, so that one can not trust the
perturbative expansion, truncated to this order.

\begin{figure}[h]\centering
\includegraphics[width=.45\textwidth]{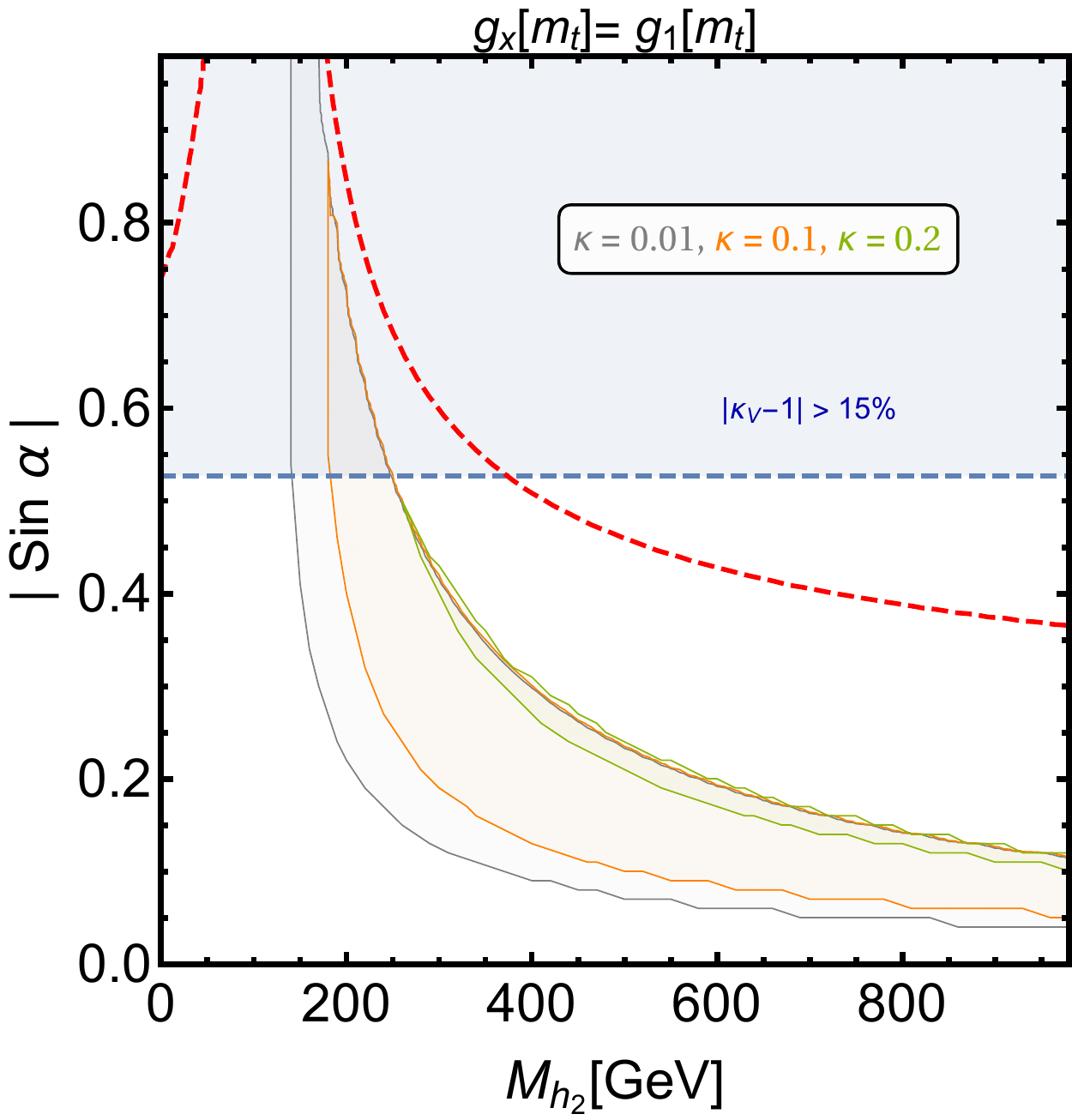}
\includegraphics[width=.45\textwidth]{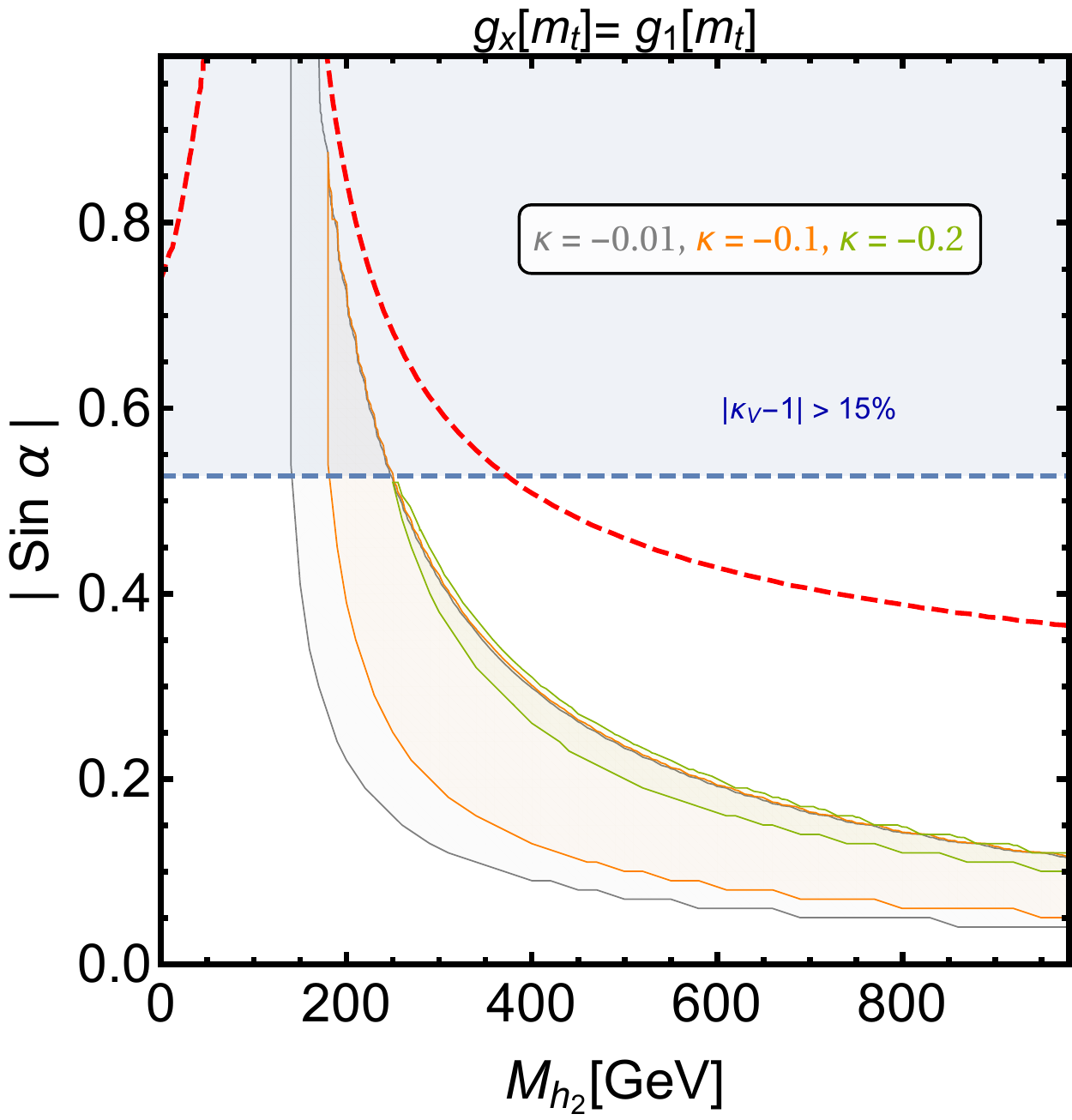}
\caption{The regions of allowed parameter space (inside the ``boomerang'' contours)  
where couplings remain perturbative, positivity is satisfied up to $M_{Pl}$ 
in the plane  $(|\sin \alpha|,\mtwo )$, at 2 loops. Left and right panels corresponds to
positive and negative $\kappa$, respectively. 
The allowed area decreases as the magnitude of $\kappa$ increases. The blue area denotes region where $h_1$ 
coupling to vector bosons is away from its SM value 
by more than 15\%. The region over dashed red line is excluded at 95\% CL by the analysis of the Peskin-Takeuchi S,T parameters.}
\label{PosCombined}
\end{figure}

The VDM model without the additional $\uone$ reduces to a Higgs portal extension of the Standard Model, 
which has been explored recently in \cite{Martin-Lozano:2015dja,Robens:2015gla,Falkowski:2015iwa}.  
These papers note that it can be useful to consider a change of parameterisation 
from $(\lambda_H,\kappa,\lambda_S) \rightarrow (\mtwo,\kappa,\sin\alpha)$, 
in which one can re-express the quartic couplings as follows:
\beq
\lamh=\lamsm+\sin^2\alpha \frac{\mtwo^2-\mone^2}{2v^2}
\eeq
\beq
\lams=\frac{2\kappa^2}{\sin^2 2\alpha}\frac{v^2}{\mtwo^2-\mone^2}\left( \frac{\mtwo^2}{\mtwo^2-\mone^2}-\sin^2\alpha  \right).
\eeq
One can then identify the regions in this parameter space,  $(|\sin\alpha|, \mtwo)$, that are absolutely 
stable (i.e. the stability conditions are satisfied up to the Planck scale)
and do not have Landau poles at any renormalisation scale before the Planck scale, as pictured in  
fig.~\ref{PosCombined}. 
We show here those plots in order to compare our results with those of \cite{Falkowski:2015iwa}, which differ 
in so far as our study include 2-loop  (not 1-loop) running, and where our model includes the 
additional parameter $g_x$. 


\section{Constrained parameter space}
\label{par_space}

In this section we show results of scans over the parameter space ($\lamh,\kappa,\lams,\gx$).
Usually $\lams$ will be fixed and specified, $\gx$ is either fixed (in figs.~\ref{megaplot1} - \ref{megaplot13})
or varied $0.1 < \gx < 1$  while $\lamh$ and $\kappa$ will be scanned such that $0<\lamh<0.25$ and
$-0.5<\kappa<0.5$ (with some exceptions when $-1<\kappa<1$). If RGE running is employed 
then the scan range should be regarded as for initial values of running quantities at $Q=\mt$. 

\subsection{Collider constraints}
\label{collider_con}

In addition to the theoretical requirements discussed above we are going to impose some experimental 
constraints. First of all there exist limits on branching ratio for invisible Higgs boson decays.
Searches for Higgs decaying invisibly has been carried out by both the
ATLAS and CMS collaborations at the LHC for various production and decay channels.
ATLAS \cite{Aad:2015uga}  considered an associated Higgs production with a vector boson 
($V = W^\pm$ or $Z$), assuming SM production they found an upper limit of 29\% at
95\% confidence level on the branching ratio of Higgs bosons decaying to invisible particles.
A search for invisible decays of Higgs bosons was also performed by CMS 
\cite{CMS:2015dia,Chatrchyan:2014tja}. Assuming Standard Model Higgs boson cross sections and acceptances, 
the observed upper limit on the invisible branching fraction was found
to be 57\% at 95\% confidence level. It turns out that within the VDM model unless the $\uone$ coupling constant $\gx$ 
is tiny or $2 \mzp > \msm\gev$ decays of the observed Higgs boson $h_1$ into $\zp\zp$ would dominate 
with branching ratio exceeding the experimental limits. Similar comments apply for decays
of $h_1$ into pairs of $h_2$ (for the scenario with $\lamh < \lamsm$). Therefore 
in our analysis we simply exclude points in the parameter space such that $h_1$ decays to $\zp\zp$ or $h_2h_2$
are kinematically allowed.  

If $\lamh<\lamsm$ Higgs boson $h_2$ is light so that it could have been prodeced at LEP. Then the LEP limits for
$e^+e^-\to Z h_2$ apply and should be imposed. Here we adopt the data from \cite{Schael:2006cr}
where limits on $ZZh_1$ coupling normalized to the SM $ZZh$ coupling ($\kappa_Z=\cos\alpha$)
are tabularised as a function of $\mtwo$. For fixed $\lams$ the limit on $\alpha$ could be
easily translated into allowed region in the ($\kappa,\lamh$) plane.

Higgs boson couplings are being measured at the LHC. The ATLAS \cite{ATLAS:Moriond_2015}
and CMS \cite{Khachatryan:2014jba} collaborations
limit e.g. ratios $\kappa_V$ of Higgs boson $VV$ couplings normalized to their SM values. The conclusion is that
the observed Higgs has SM-like couplings to the vector bosons. Here we will assume, somehow arbitrarily, that
the ratio is limited by $0.85 < \kappa_V <1$. Note that because of the orthogonal mixing $\kappa_V=\cos\alpha$
can not exceed $1$ within the VDM model. As we have already mentioned above the SM-like nature of $h_1$
favours regions of large and small $\vx$ for $\lamh>\lamsm$ and $\lamh<\lamsm$, respectively.
The constraint that originates from $0.85 < \kappa_V <1$ could be expressed as an allowed region
in the $(\kappa,\lamh)$ plane both for $\lamh<\lamsm$ and $\lamh>\lamsm$ case. 

 In order to estimate relevance of electroweak precision data we adopt the Peskin-Takeuchi 
S, T and U parameters \cite{Peskin:1991sw}\footnote{Since the scale of new physics in our case is not always much above the electroweak scale
therefore  the S, T and U parameters should be used with some extra care. In particular, for $\mtwo<\mone$ they do not
provide a viable estimation of radiative corrections. Luckily that region of parameter space is not allowed by other constraints.}.
At the 1-loop level, beyond the SM radiative corrections to the SM vector boson progators $\delta\Pi_{VV}$ are not affected by the presence of the $Z'$ boson, 
therefore $\delta\Pi_{VV}$ are the same as those found in the analysis of the plain Higgs portal \cite{Falkowski:2015iwa}. 
At this order shifts in $\delta\Pi_{Z\gamma}$ and $\delta\Pi_{\gamma\gamma}$ vanish, therefore the S and T parameters can be expressed as 
\beq
S=\frac{16 \pi \cos^2 \theta_W}{g^2} \delta\Pi'_{ZZ}(0),\;\;\;\;\;T=\frac{4\pi}{e^2}\left(\frac{\delta\Pi_{WW}(0)}{M^2_W}-\frac{\delta\Pi_{ZZ}(0)}{M^2_Z}\right),
\eeq
whereas the parameter $U$  is too small to be relevant. Using the fit obtained in \cite{Ciuchini:2014dea} we found that at 95\% confidence 
level the S and T parameters do not constrain further the parameter space. These bounds are entirely embedded in region, where the scalar 
mixing angle is too large or couplings are non-perturbative (see fig.~\ref{PosCombined}). It is worth to emphasize that
they constrain $|\sin\alpha|$ only moderately weaker than the full set of the electroweak precision observables \cite{Falkowski:2015iwa}. 
In particular, in the important mass range $\mtwo > 200$~GeV, the allowed value of $|\sin\alpha|$ in $S$ and $T$ approximation is 
larger by $25\%$. For low $\mtwo$ mass differences grow however that region of parameter space is anyway disfavoured by the requirement of 
absolute stability and excluded by limits on $\kappa_V$.

\subsection{Dark Matter abundance}
\label{abundance}

\begin{figure}[h]\centering
\includegraphics[width=.45\textwidth]{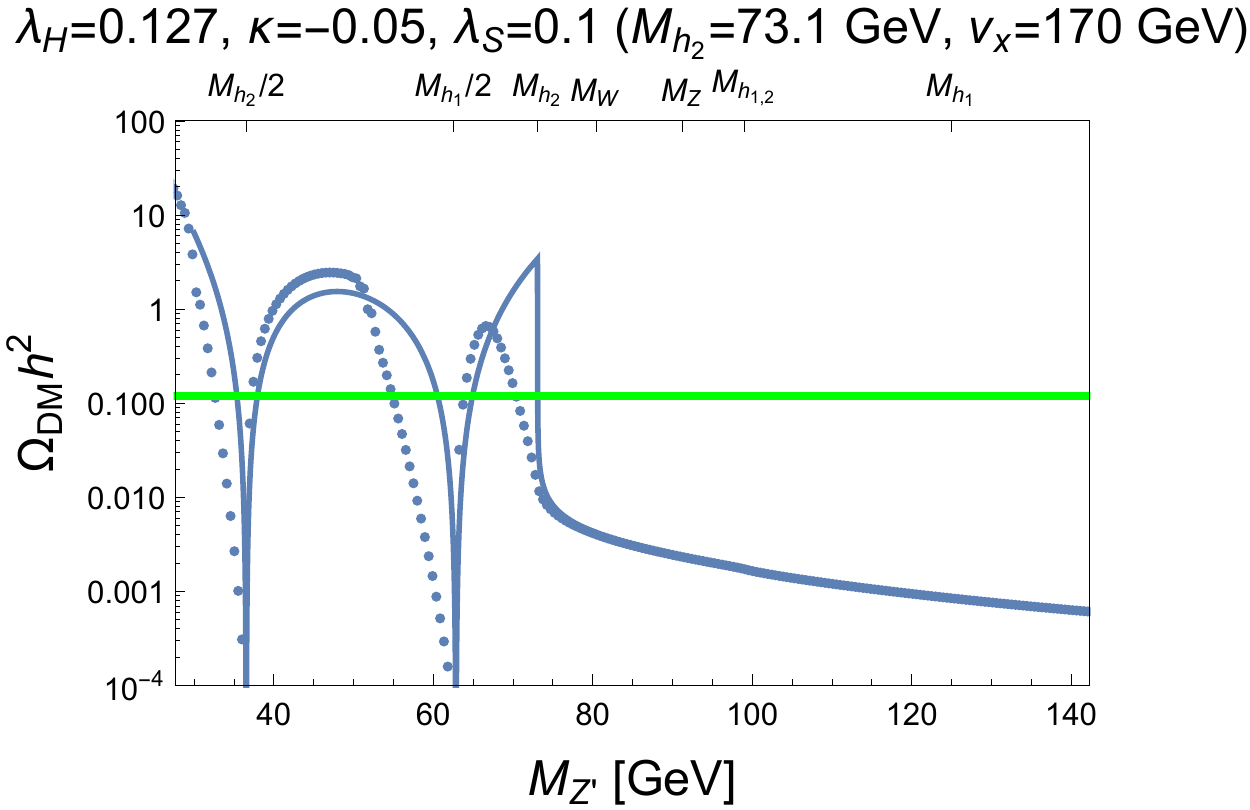}
\includegraphics[width=.45\textwidth]{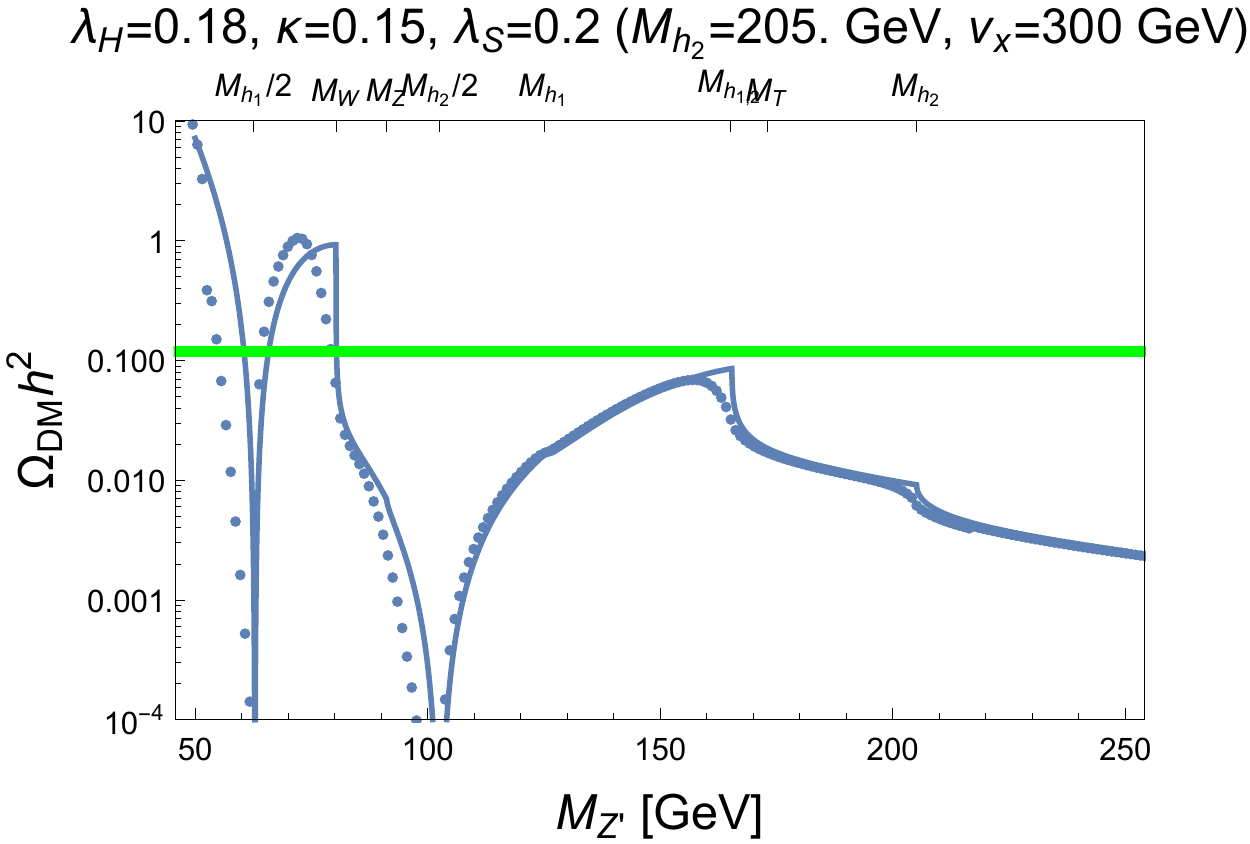}\\
\includegraphics[width=.45\textwidth]{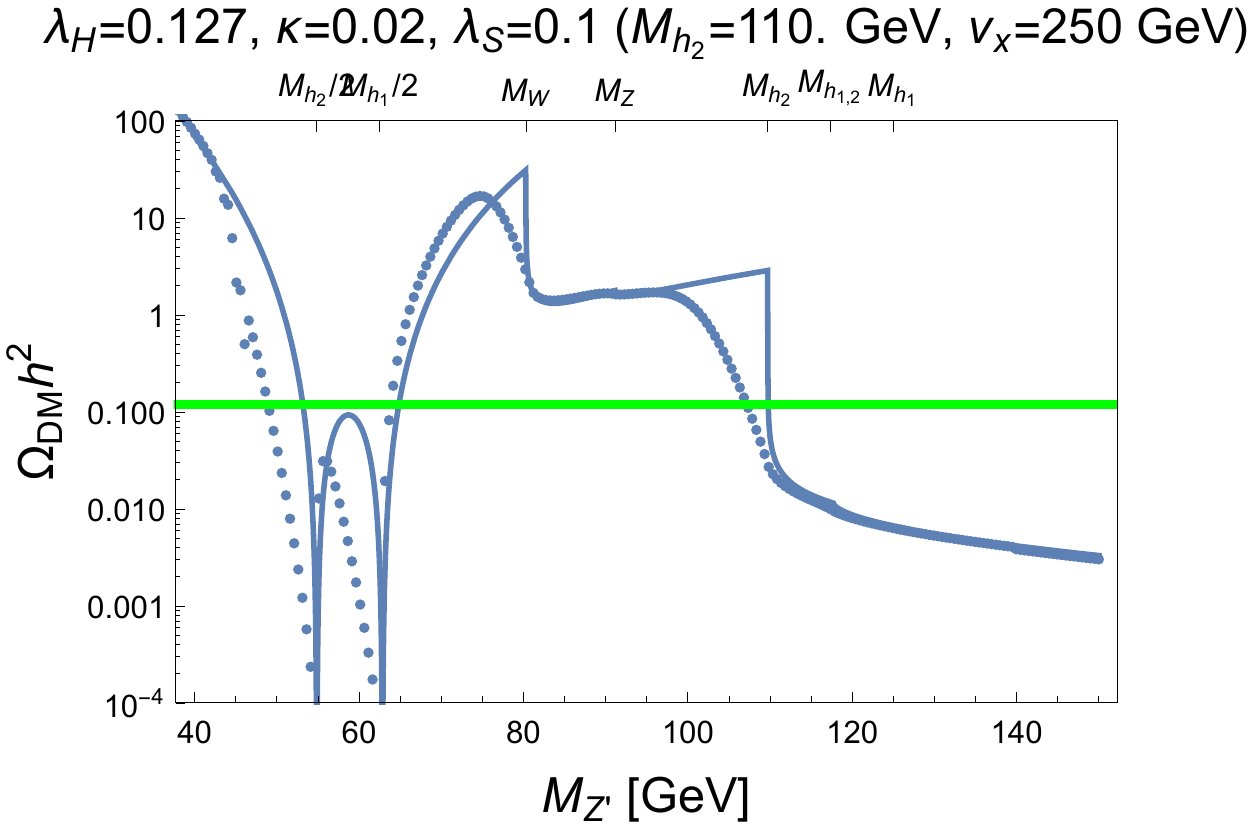}
\includegraphics[width=.45\textwidth]{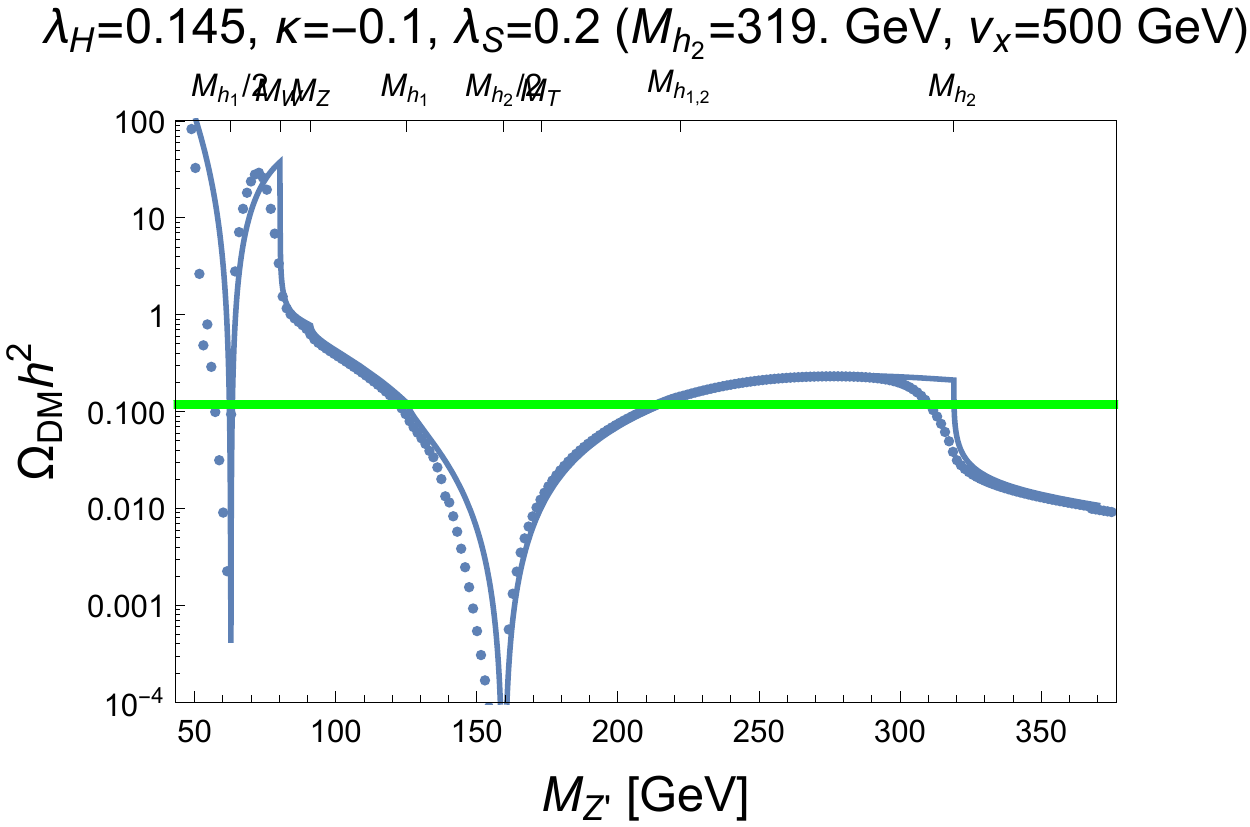}\\
\caption{The left and right panels show $\odm$ as a function of $\mzp$ 
for a case of $\mtwo < \msm\gev$ ($\lamh<\lamsm$) and $\mtwo > \msm\gev$ ($\lamh>\lamsm$),
respectively. Masses of $\zp$ that correspond to a resonance or an opening of a new final states
are shown above each panel. The dotted and solid lines correspond to results obtained from 
the micrOMEGAs and results of cold dark matter approximation from the 
appendix~\ref{app:rel_abun}, respectively.
The thick horizontal green lines correspond to the observed
$5\sigma$ result for $\odm$ as in (\ref{Omega_cmb}).}
\label{omegaplot}
\end{figure}

Before we proceed to constrain the parameter space by measurements of DM abundance,
in fig.~\ref{omegaplot} we show results for $\Omega_{DM} h^2$ as a function of the DM mass ($\mzp$)
obtained varying coupling constant $g_x$ and choosing a few representative values of other parameters.

We have calculated $\Omega_{DM} h^2$ adopting standard textbook methods for cold relics, see e.g.
\cite{Kolb:1990vq}. Relevant vertices, Feynman diagrams and corresponding contributions to 
$\zp\zp$ annihilation cross section for various final states are collected in appendix~\ref{app:rel_abun}.
We have checked our results for $\Omega_{DM} h^2$ against calculations done adopting 
the micrOMEGAs3~\cite{Belanger:2013oya}. It turned out that except resonance regions (such that
$2\mzp\sim$ a mass of s-channel resonance) and vicinities of 
thresholds for new final states, $\Omega_{DM} h^2$ determined via the cold relics technique
agrees pretty well with the result provided by the micrOMEGAs. However, in order to have also those special regions under control 
we have decided to adopt the micrOMEGAs hereafter. For illustration, in \ref{omegaplot} 
we compare $\Omega_{DM} h^2$ obtained by adopting results contained 
in appendix~\ref{app:rel_abun} with those from micrOMEGAs.

There is a comment here in order. In fig.~\ref{omegaplot} and similar that will follow, one sees that for large $\mzp$ it is typical
that $\odm$ decreases as $\mzp$ grows.
In fact it is easy to understand such behaviour. In those plots potential parameters are fixed, so is $\vx$, therefore increasing $\mzp$ implies 
growing $\gx$, so that $\langle \sigma v\rangle$ increases and therefore $\odm$ decreases. This fact can be seen in the easiest way by looking 
at the contribution from direct DM-scalar interaction coming from the vertex $V^{Z'}_{ij}\propto g_x^2$ (appendix A); 
then $\langle \sigma v\rangle \propto |V^{Z'}_{ij}|^2/M_{Z'}^2\propto g_x^2/v_x^2$ and therefore $\odm\propto 1/\gx^2$.

\subsection{Allowed parameter space}
\label{allowed_par_space}

\begin{figure}[h]\centering
\includegraphics[width=.9\textwidth]{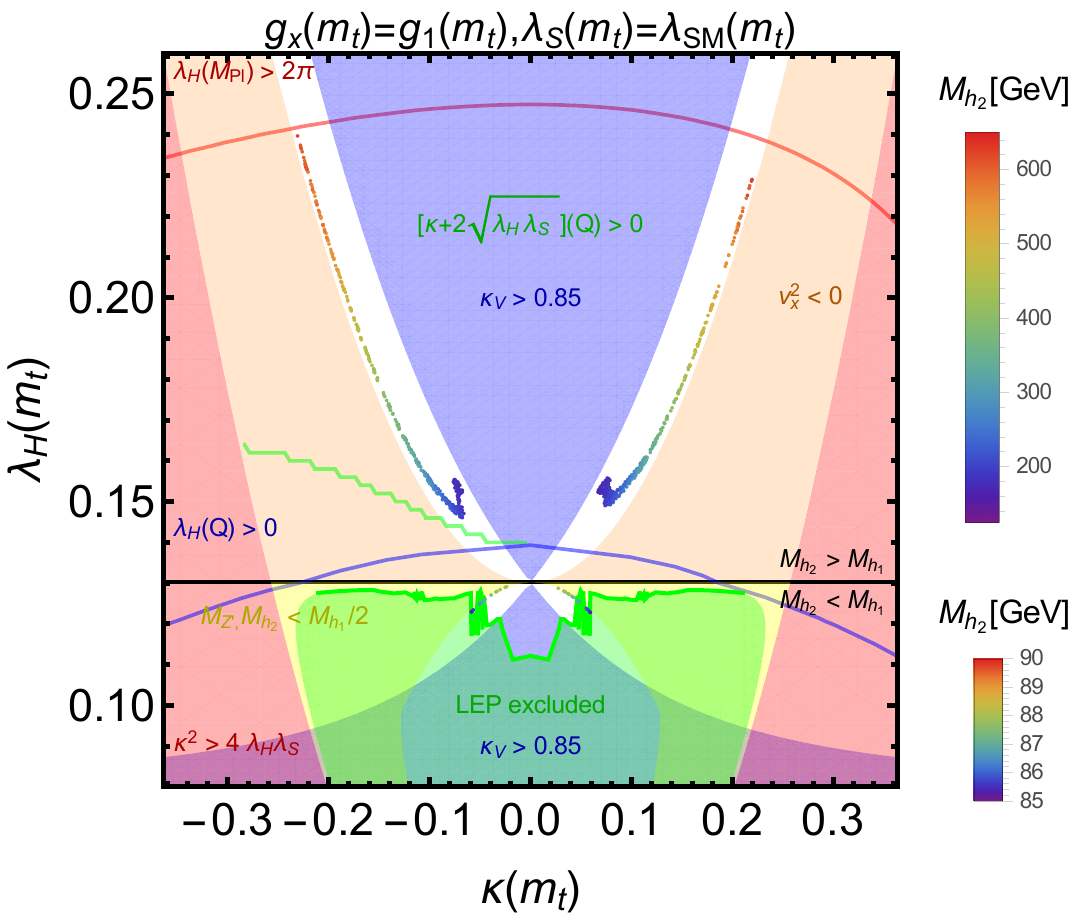}
\caption{Combined plots of allowed and disallowed parameter space in the plane $(\lambda_H(\mt),\kappa(\mt) )$
for $\gx(\mt)=g_1(\mt)$ and $\lams(\mt)=\lamsm(\mt)=0.13$. 
The thin red line denotes the frontier above which a Landau pole of $\lambda_H$ appears below the Planck scale. 
The thin blue line denotes the absolute stability frontier.  Below the thin green line the  positivity 
condition fails at some renormalisation scale (its wavy shape is a numerical artifact). 
The yellow region is disallowed by invisible Higgs decays.
The green area denotes LEP exclusions on Higgs-like scalars. 
In the outer red area positivity fails at the low scale, while in the orange area no physical solution 
of the vev $\vx$ exists. The blue area denotes an excess of the $h_1$ Higgs couplings to vector bosons $(\kappa_V)$. 
The remaining allowed region is in white. The points (in the white region) coloured with respect to the mass of the extra Higgs boson $\mtwo$ 
are those for which also $\odm$ constraint (\ref{Omega_cmb}) is fulfilled.}
\label{megaplot1}
\end{figure}

\begin{figure}[h]\centering
\includegraphics[width=.9\textwidth]{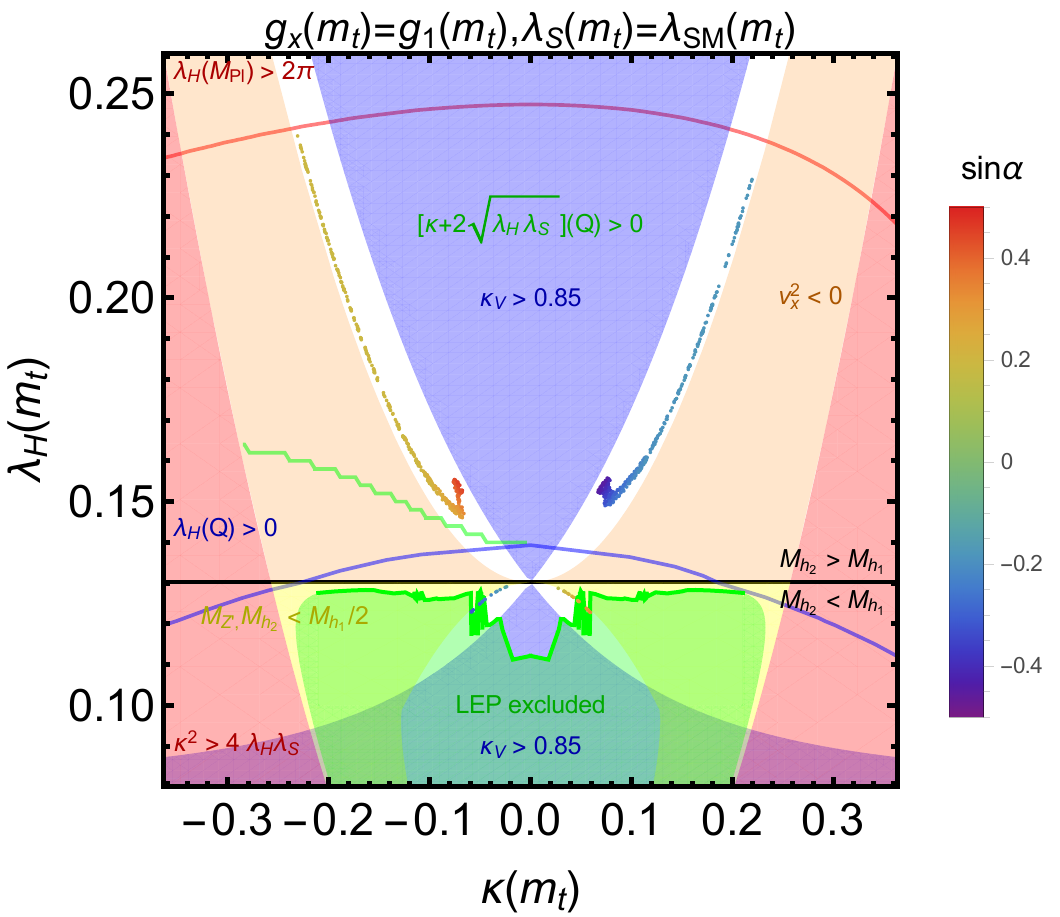}
\caption{Same as in fig.~\ref{megaplot1} however colouring of allowed points is here with respect to $\sin\alpha$.
}
\label{megaplot2}
\end{figure}

\begin{figure}[h]\centering
\includegraphics[width=.452\textwidth]{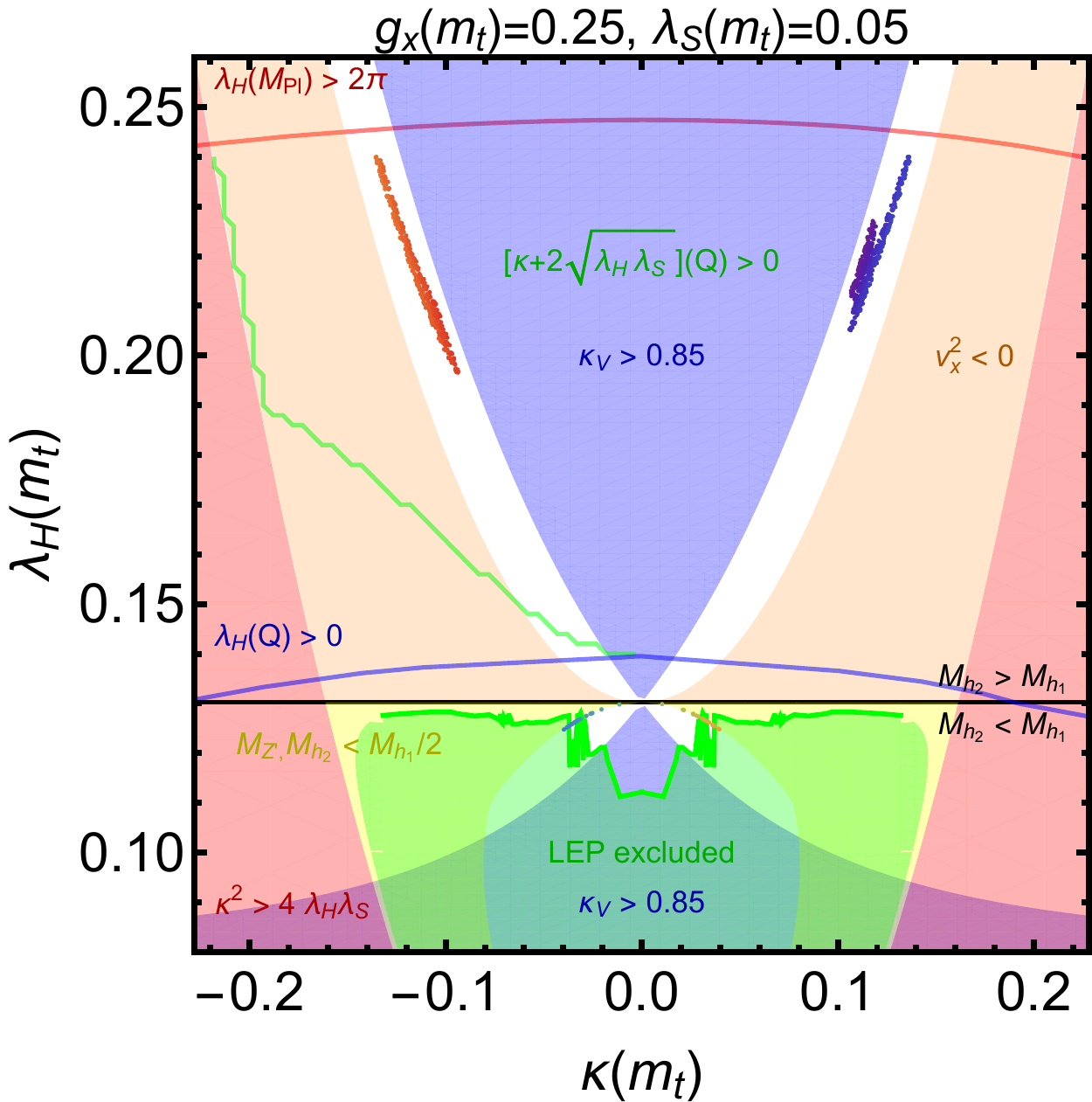}
\includegraphics[width=.537\textwidth]{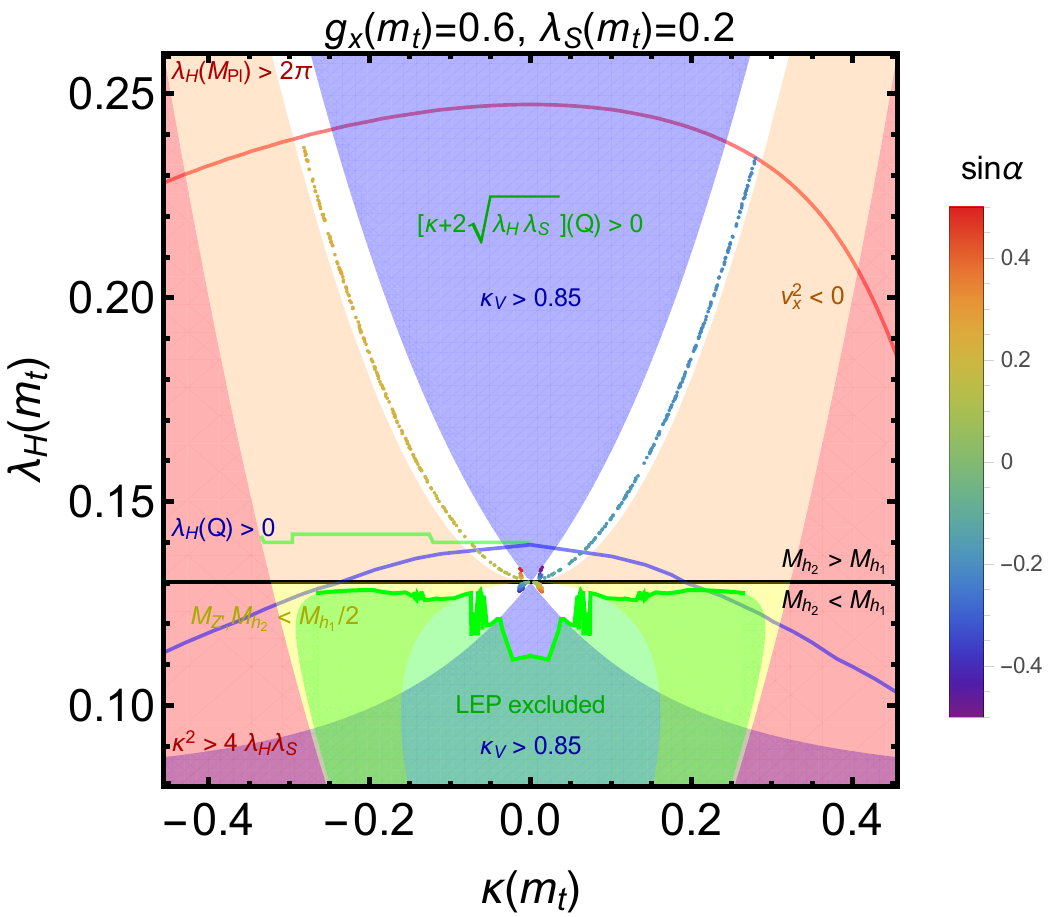}
\caption{Same as in figs.~\ref{megaplot2}, however for $\gx(\mt)=0.25$ and $\lams(\mt)=0.05$
(left panel) and for $\gx(\mt)=0.6$ and $\lams(\mt)=0.2$ (right panel).}
\label{megaplot13}
\end{figure}

All constraints that we have considered are collected in figs.~\ref{megaplot1}, \ref{megaplot2} and \ref{megaplot13}.
Choosing randomly points in the parameter space in the region (white) allowed by perturbativity, stability, LEP and LHC data
we require that the DM abundance remains within the $5\sigma$ limit of \cite{Ade:2015xua}
\beq
\Omega_{DM} h^2=0.1199\pm 0.0022
\label{Omega_cmb}
\eeq
Points that fit into the allowed region of (\ref{Omega_cmb}) are shown in figs.~\ref{megaplot1}-\ref{megaplot13} 
as dots within the white region, for them all the constraints are satisfied.

\subsection{Direct detection of Dark Matter}
\label{dd}

Interactions of VDM with nucleons are mediated by the Higgs particles. For the elastic scattering cross section we obtain
\beq
\xdd=\frac{\mu^2}{4\pi}\gx^2 g_{hNN}^2 \sin^2 2 \alpha \left(\frac{1}{M_{h_1}^2}-\frac{1}{M_{h_2}^2}\right)^2,
\eeq
where $\mu=\mzp M_N/(M_N+\mzp)$ is nucleon-DM reduced mass, $g_{hNN}=1/v \langle N | \sum_q m_q \bar{q}q | N \rangle $ 
is the effective Higgs to nucleon coupling \cite{Cheng2014}.

In this subsection we are going to show results for $\xdd$
as a function of $\mzp$ for those points (green points in figs.~\ref{megaplot1} - \ref{megaplot13}) 
in the parameter space which satisfy all the constraints including the DM abundance. 
In the following subsection we are showing results obtained with \MICRO for $\lams$ fixed at $\lams=0.2$.

\subsubsection{Light dark matter}

\begin{figure}[h!]\centering
\includegraphics[width=0.8\textwidth]{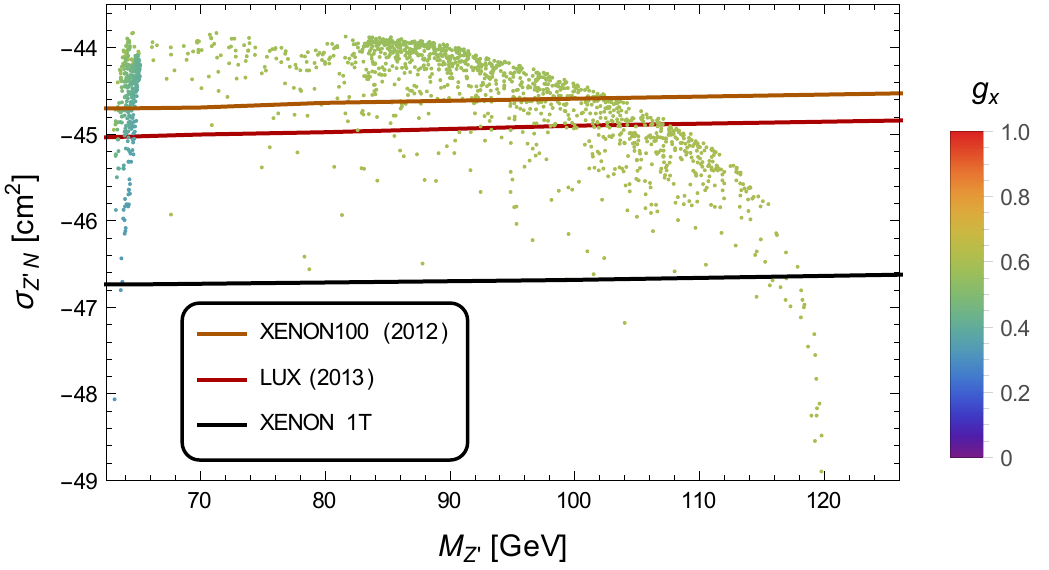}
\caption{The figure shows the DM-nucleon cross section, $\xdd$, as a function of
the DM mass $\mzp$ for points which satisfy all other constraints for $\lamh<\lamsm$. The singlet
quartic coupling is fixed at $\lams=0.2$. Colouring corresponds to the strength of the gauge coupling $\gx$.
The nearly horizontal lines are the experimental limits for $\xdd$ from XENON100,
LUX (2103) and anticipated results for XENON 1T.}
\label{ConLow}
\end{figure}

\begin{figure}[h!]\centering
{\includegraphics[width=.49\textwidth,height=0.22\textheight]{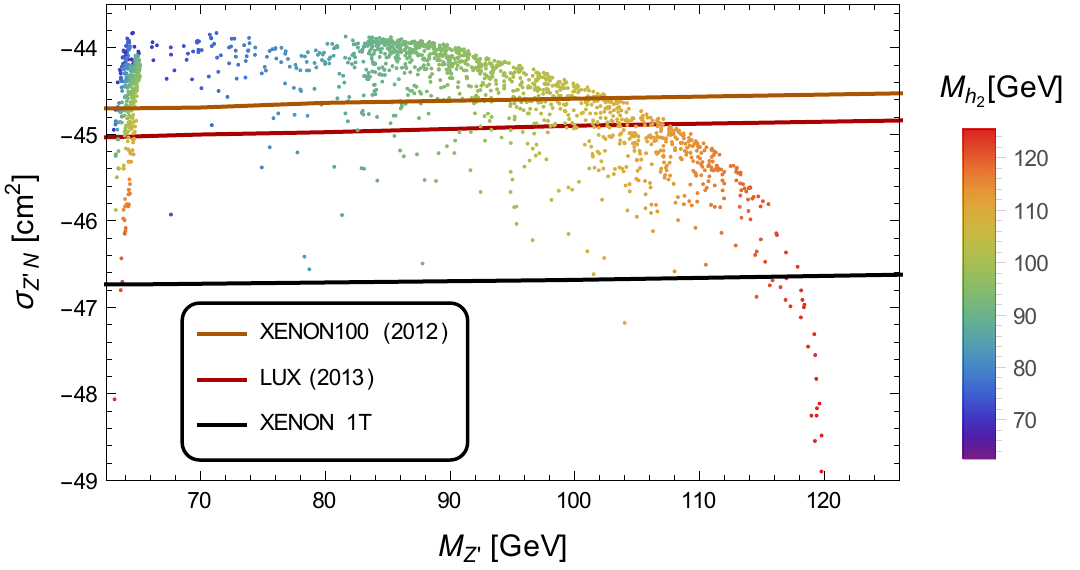}}
{\includegraphics[width=.49\textwidth,height=0.22\textheight]{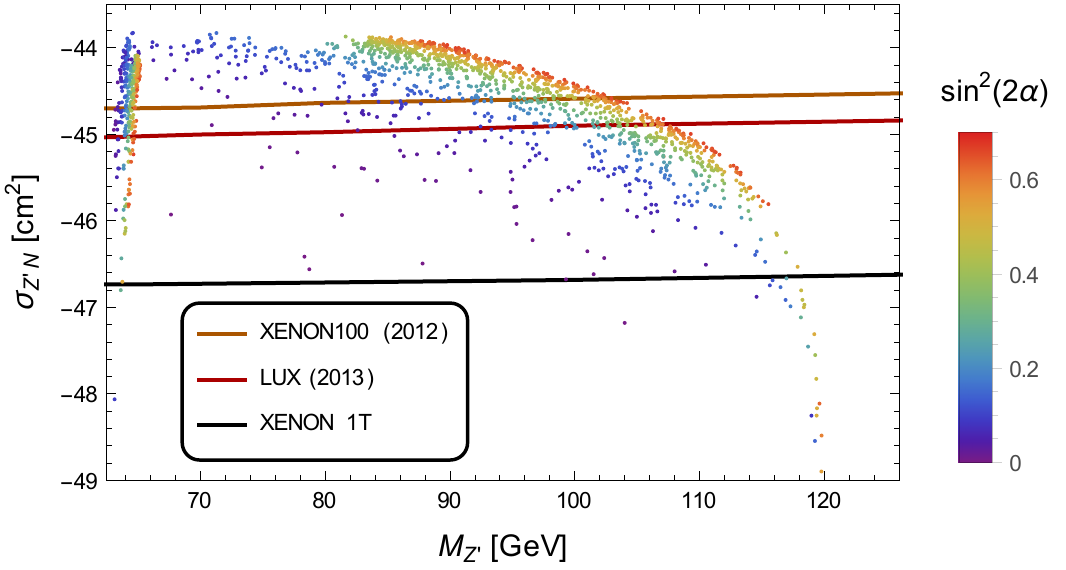}}
\caption{Same as in fig.~\ref{ConLow}, but colouring is with respect to $\mtwo$ and 
$\sin^2(2\alpha)$ for the left and the right panel, respectively. }
\label{DD-massL}
\end{figure}

We start with the case of $\lamh < \lamsm$. Then, since $\vx$ is limited from above
as in (\ref{vx_lim_down}) and the scanning interval for $\gx$ is $0.1<\gx<1$,
therefore the DM mass is bounded from above by $\mzp < v (\lamsm/\lams)^{1/2}\sim 200\gev $.
It turns out that in practice there is no consistent points in the parameter space with  
$\mzp \geq 120\gev$.  
Note also that in order to prevent invisible  $h_1$ decays $h_1 \to \zp \zp$ it is required
that $\mzp > \msmhalf\gev$, so the allowed range for DM mass is $\msmhalf\gev < \mzp < 120\gev$. 
In figs.~\ref{ConLow} and \ref{DD-massL} we are showing $\xdd$
coloured with respect $\gx$, $\mtwo$ and $\sin^2(2\alpha)$ in order to learn properties 
of the points that are plotted. We also show experimental limits  for $\xdd$ from XENON100,
LUX (2013) and anticipated results for XENON 1T. As it is seen from the figures, 
points that are consistent with the present data correspond to medium gauge coupling
$\gx \sim 0.5$, $h_2$ slightly lighter than the observed Higgs $\mtwo \sim 110 \div 125\gev$ and small mixing angle $0 \lsim \alpha \lsim \pi/8$. The DM mass
varies between $60$ and $120\gev$ with heavier states favoured. 

\begin{figure}[h!]\centering
{\raisebox{0.0\height}{\includegraphics[width=.49\textwidth]{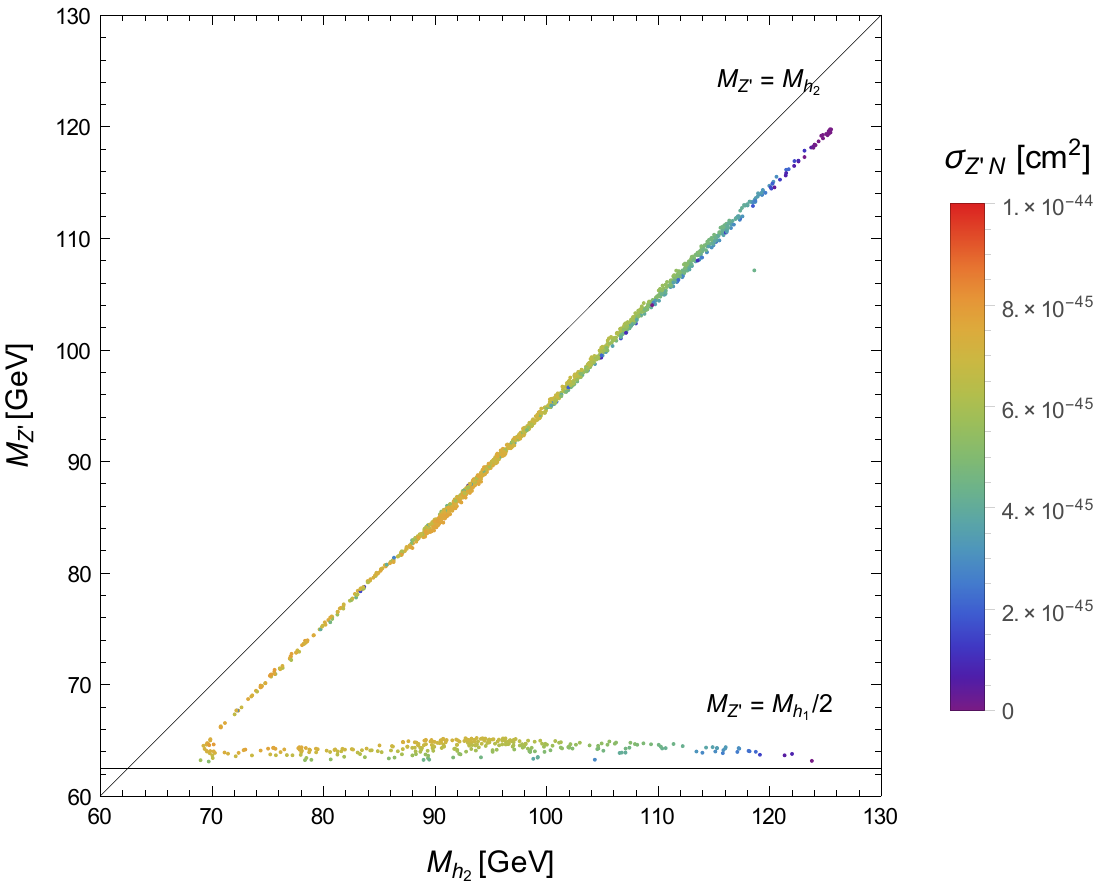}}}
\includegraphics[width=.49\textwidth,height=0.26\textheight]{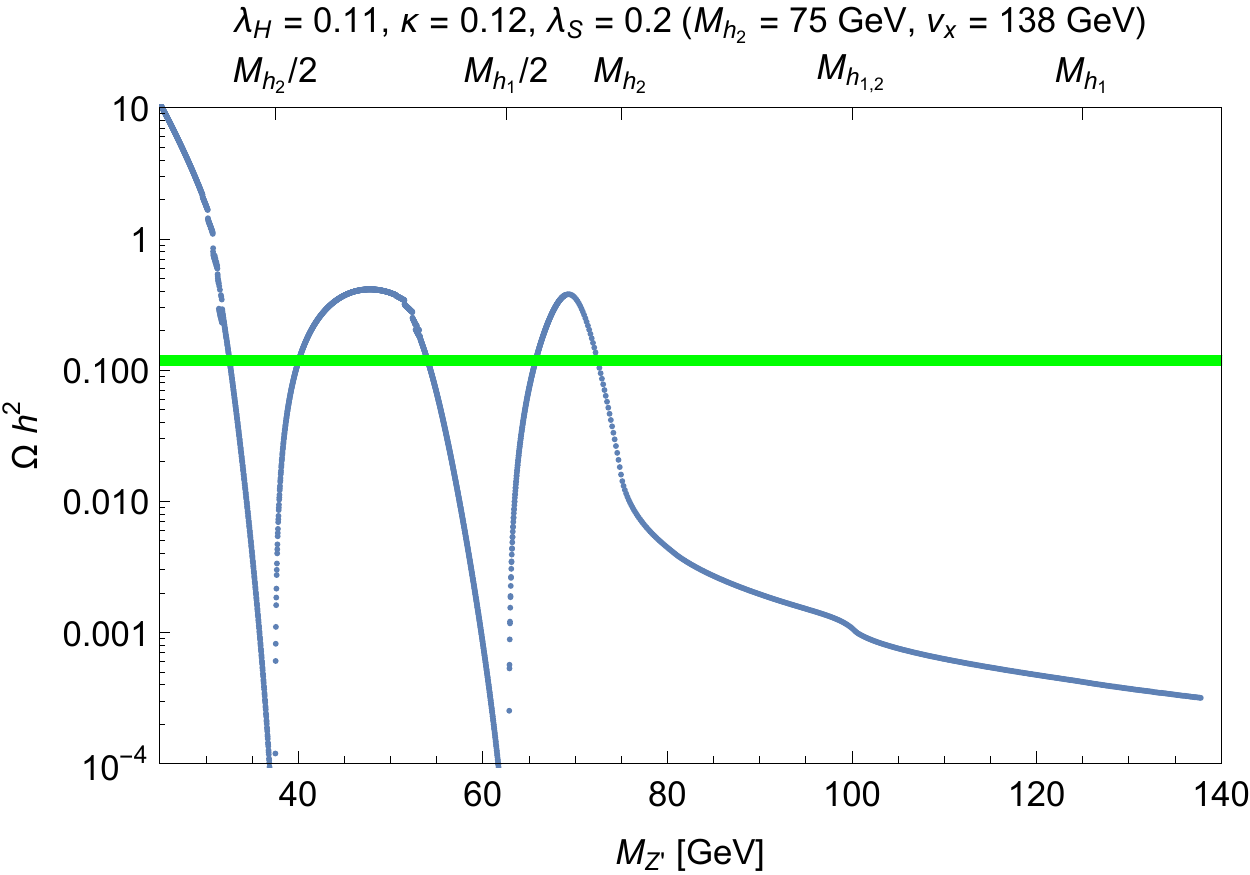}
\caption{The left panel illustrates correlation between between $\mtwo$ and $\mzp$,
while the right one shows predictions for $\odm$ as a function of $\mzp$. The colouring
corresponds to the cross section $\xdd$. Above the
right box resonances and channels which open as $\mzp$ increases are shown.
Coordinates in the parameter space ($\lamh,\kappa,\lams$) and corresponding
$\mtwo$ and $\vx$ are shown above the right panel.}
\label{DM-N_xsec_1}
\end{figure}

Fig.~\ref{DM-N_xsec_1} is helpful in order to understand why there is no 
points in the parameter space with $\mzp \gsim 120\gev$. As it is seen in the 
right panel of the figure for large $\mzp$ the DM abundance $\odm$ is very small since almost 
all annihilation channels (except $h_1 h_1$ and $t\bar t$) 
are open for $\zp$ of that mass (so that the annihilation cross section is large).
It is also instructive to look at correlations between $\mtwo$ and $\mzp$ shown
in the left panel of fig.~\ref{DM-N_xsec_1}. It turns out that there are
two regions consistent with the constraints: {\it i)} $\mzp \sim 65\gev$
and {\it ii)} $\mzp \sim \mtwo - 5\gev$.  The first one corresponds to the vicinity
of $h_1$ resonance in the right plot in the figure. The dip at $\msmhalf\gev$ is
quite steep  such that deviation by about $5\gev$ is sufficient to reach $\odm \sim 0.1$.
The other side of the resonance is excluded by the requirement of
no invisible $h_1$ decays. In fact it also easy to understand the second region
that corresponds to the other side of the summit seen to the right of the $h_1$
resonance in the right panel. There the sudden drop of $\odm$ is caused by the opening 
of $h_2 h_2$ final state. In a real case, since annihilating $\zp$ pairs are not 
exactly at rest, therefore the $h_2 h_2$ channel opens by $5\gev$ earlier.


\subsubsection{Heavy dark matter}

\begin{figure}[h]\centering
\includegraphics[width=0.8\textwidth]{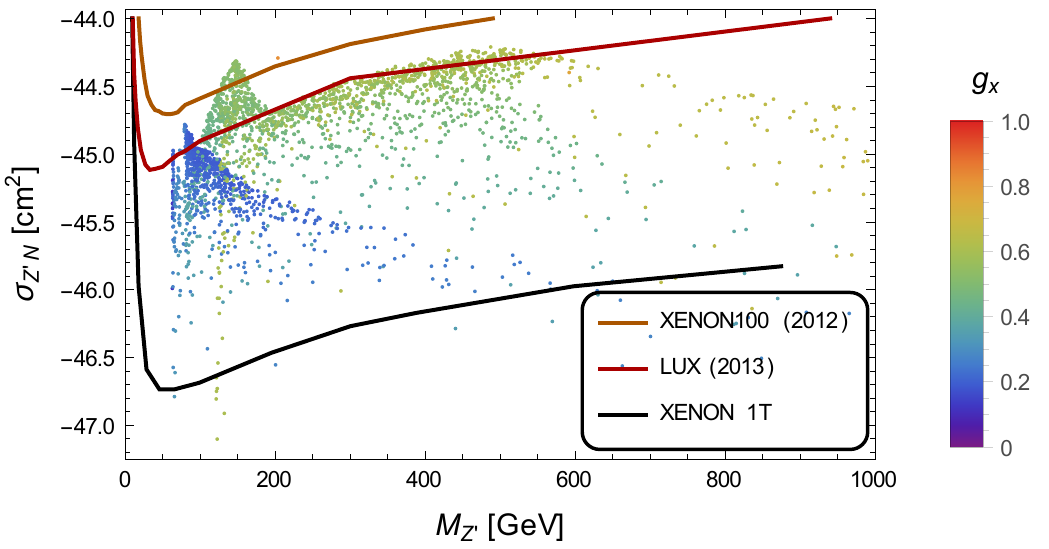}
\caption{The figure shows the DM-nucleon cross section, $\xdd$, as a function of
the DM mass $\mzp$ for points which satisfy all other constraints for $\lamh>\lamsm$. The singlet
quartic coupling is fixed at $\lams=0.2$. Colouring corresponds to the strength of the gauge coupling $\gx$.
The solid lines are the experimental limits for $\xdd$ from XENON100,
LUX (2103) and anticipated results for XENON 1T.}
\label{ConUp}
\end{figure}

\begin{figure}[h]\centering
{\includegraphics[width=.49\textwidth,height=0.22\textheight]{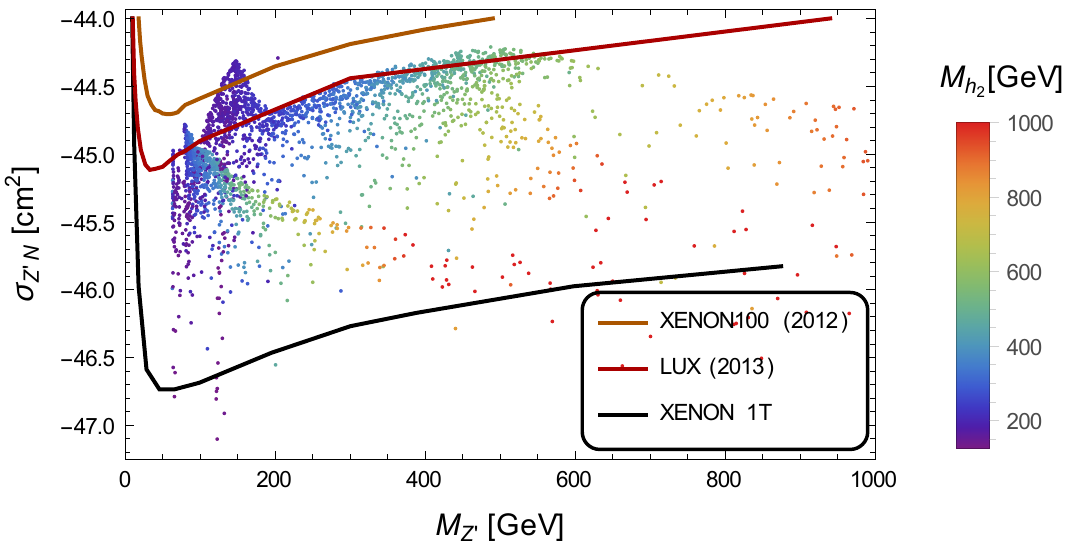}}
{\includegraphics[width=.49\textwidth,height=0.22\textheight]{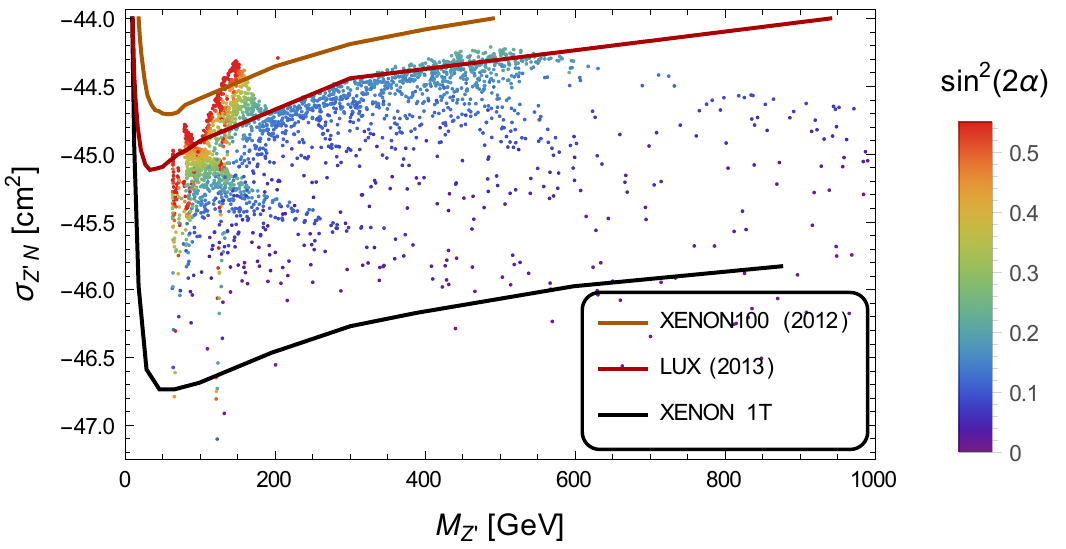}}
\caption{Same as in fig.~\ref{ConUp}, but colouring is with respect to $\mtwo$ and 
$\sin^2(2\alpha)$ for the left and the right panel, respectively. }
\label{DD-mass}
\end{figure}

In this section we are considering the case $\lamh > \lamsm$.
Then, since $\vx$ is limited from below as in (\ref{vx_lim_up}) and the scanning 
interval for $\gx$ starts at $0.1$ therefore the DM mass is bounded from below by 
$\mzp >  0.1 v (\lamsm/\lams)^{1/2}\sim 20\gev $. However since invisible Higgs 
decays should be prevented therefore $\zp$ must be even heavier, so that here
$\mzp > \msmhalf\gev$. Results for $\xdd$ are presented in a similar manner as in
the case of the light DM, so in figs.~\ref{ConUp} and \ref{DD-mass} we are showing
the cross section plotted against the DM mass $\mzp$ with colouring corresponding to
$\gx$, $\mtwo$ and $\sin^2(2\alpha)$. As it is seen from the figures there exist
points that lay below the LUX 2013 upper limit, they correspond to medium gauge coupling
$\gx \sim 0.2 \div 0.6$, wide range of $\mtwo$ varying from $130\gev$ up 
to $1000\gev$ and  small mixing angle $0 \lsim \alpha \lsim \pi/8$. The allowed DM mass 
varies between $\msmhalf\gev$~to~$1000\gev$. 

\begin{figure}[h]\centering
{\raisebox{-0.0\height}{\includegraphics[width=.49\textwidth]{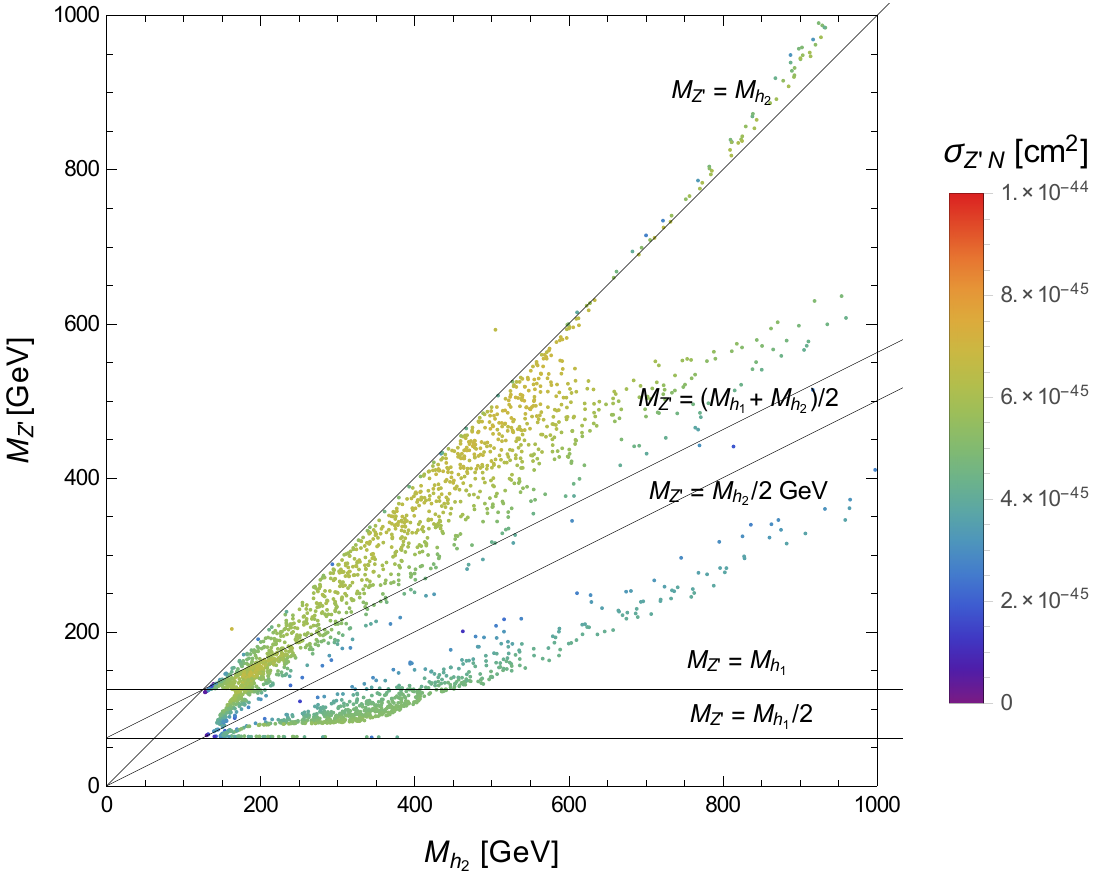}}}
{\includegraphics[width=.49\textwidth,height=0.26\textheight]{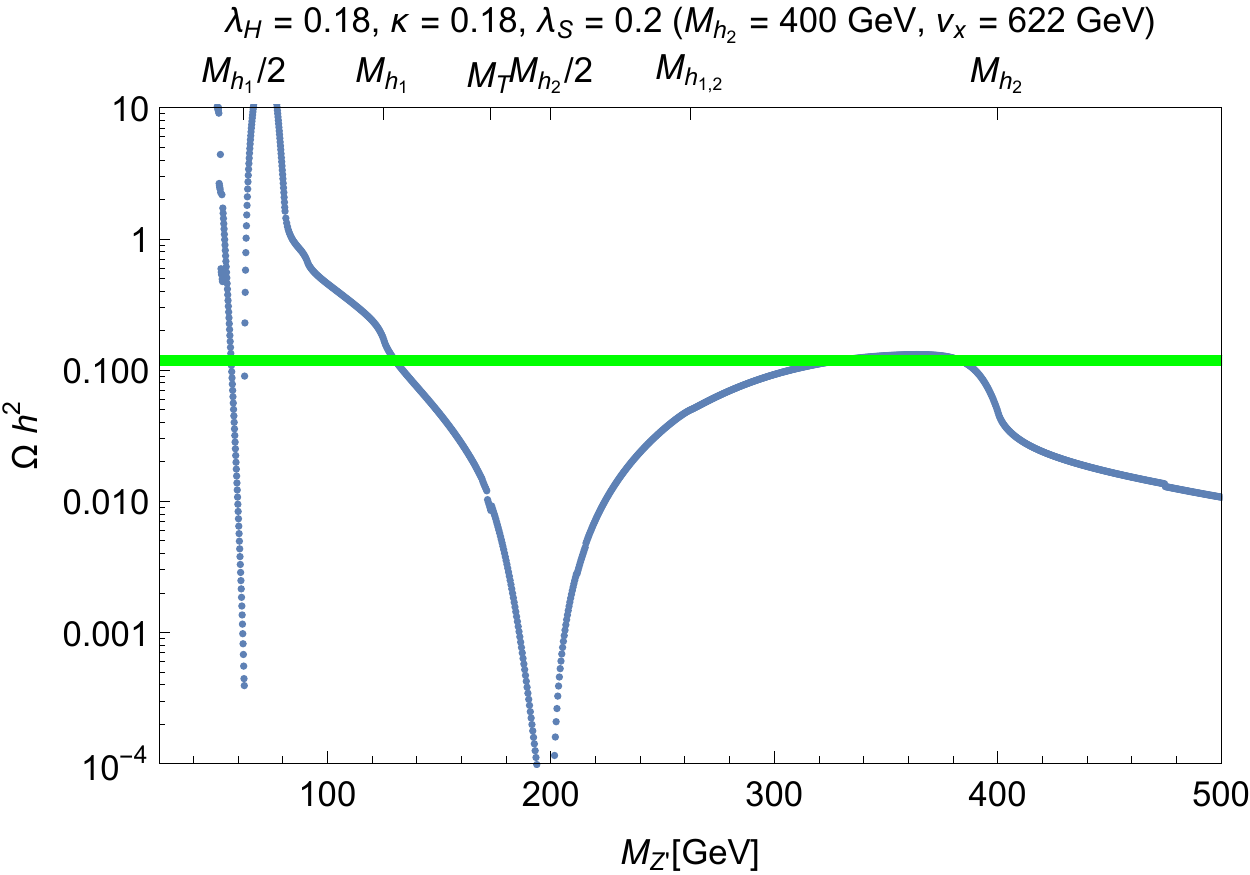}}
\caption{The left panel illustrates correlation between between $\mtwo$ and $\mzp$,
while the right one shows predictions for $\odm$ as a function of $\mzp$. The colouring
corresponds the the coupling $\xdd$. Above the
right box resonances and channels which open as $\mzp$ increases are shown.
Coordinates in the parameter space ($\lamh,\kappa,\lams$) and corresponding
$\mtwo$ and $\vx$ are shown above the right panel.}
\label{DM-N_xsec_2}
\end{figure}

\begin{figure}[h]\centering
{{\includegraphics[width=.49\textwidth]{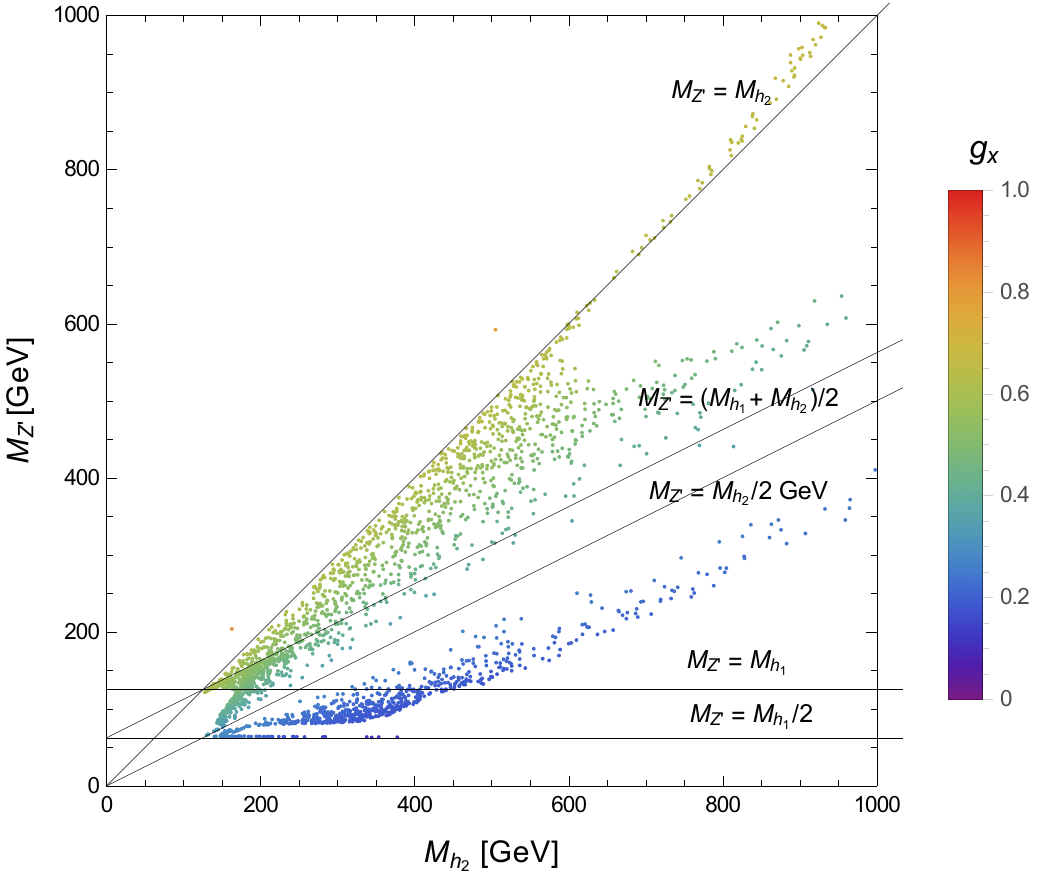}}}
{\includegraphics[width=.49\textwidth,height=0.26\textheight]{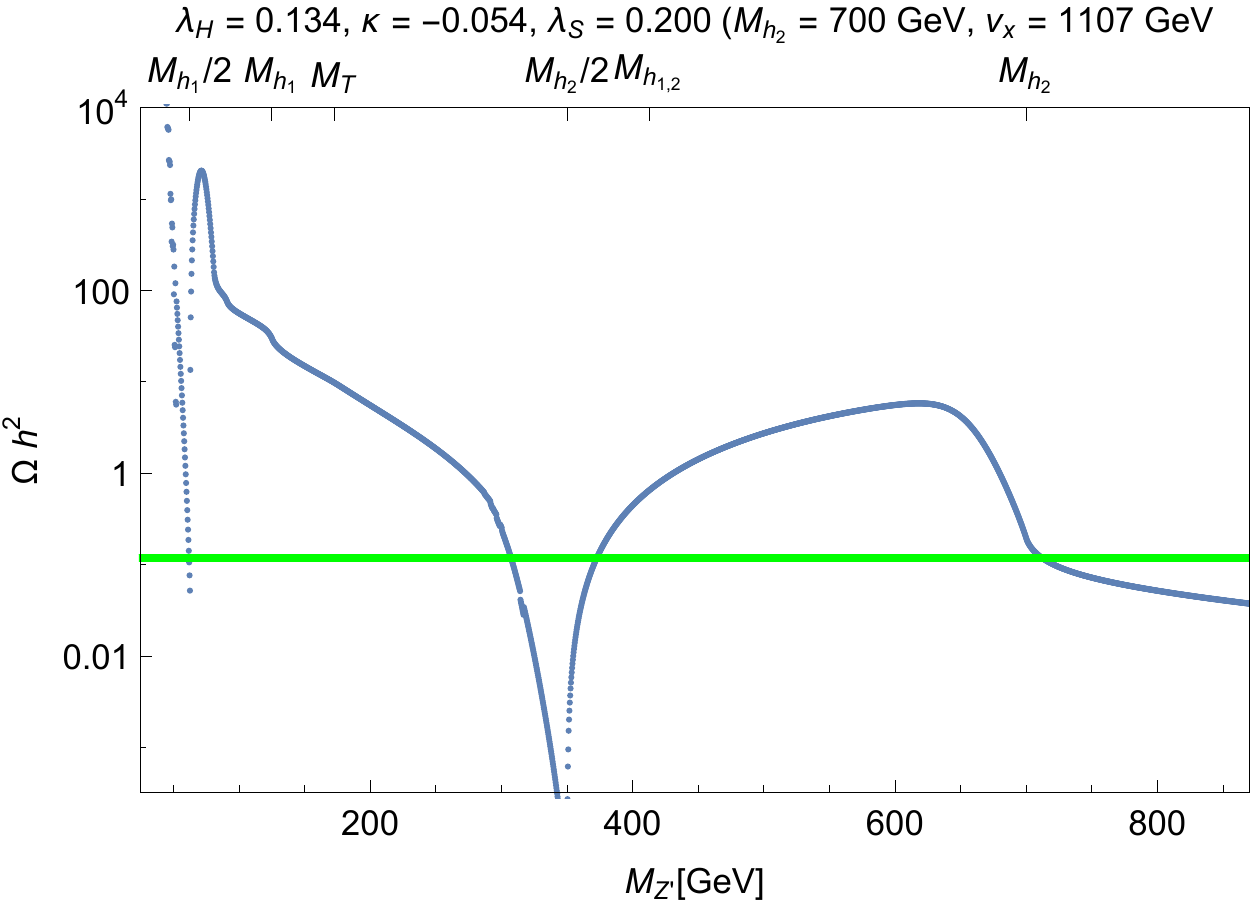}}
\caption{The left panel illustrates correlation between between $\mtwo$ and $\mzp$,
while the right one shows predictions for $\odm$ as a function of $\mzp$. The colouring
corresponds to the cross section $\gx$. Above the
right box resonances and channels which open as $\mzp$ increases are shown.
Coordinates in the parameter space ($\lamh,\kappa,\lams$) and corresponding
$\mtwo$ and $\vx$ are shown above the right panel.}
\label{relicUpper700}
\end{figure}

It is worth understanding
the origin of points that compose the blue and green hills in the vicinity of $100\gev$
in fig.~\ref{ConUp}.
As seen from the figure they correspond to different strengths of the gauge coupling 
constant: the left one is made of points with $\gx\sim 0.2$ while for the
right one the coupling is typically larger $\gx\sim 0.5$. The purpose of right panels in
figs.~\ref{DM-N_xsec_2}
and \ref{relicUpper700} is to illustrate $\odm$ dependance on $\mzp$ for 
representative points $(\lamh,\kappa,\lams)$ in the
parameter space. Note in the left panels of the figures that points in 
the space  $(\mzp,\mtwo)$ are grouped into two rough sets, one  for
$\mzp \gsim \mtwo/2$ and the other for $\mzp \lsim \mtwo/2$. Inspecting 
the right panel of fig.~\ref{DM-N_xsec_2} it is easy to see that the later group is
composed of points sitting on the left slope of a heavy Higgs ($h_2$) resonance (at
$\mtwo = 400\gev$ for this particular example). Since the resonance dip is relatively wide
in this case, therefore the shift above the the nominal resonance is substantial, typically of 
the order of $100\gev$. Then when $\mzp$ increases eventually $h_2 h_2$ channel opens up,
the cross section increases and $\odm$ drops leading to higher $\mzp$ consistent with
the observed DM abundance, that explains the other group of points in e.g.~fig.\ref{DM-N_xsec_2}
located above $\mzp \sim \mtwo/2$. The two groups are separated by the presence of the
summit in $\odm$ between them. The fig.~\ref{relicUpper700} shows that the two groups
have different $\gx$ and therefore they could be identified as the points that
compose the two hills in fig.~\ref{ConUp}: the blue hill is made of points
that lay below $\mzp \sim \mtwo/2$ while the green one of those with 
$\mzp \gsim \mtwo/2$. The right panel in fig.~\ref{relicUpper700} illustrates the mechanism
of the very strong correlation between $\mzp$ and $\mtwo$ that is observed for
large $\mzp$ along the line $\mzp \sim \mtwo$. As it is seen from the figure
the correlation is caused by the steep drop in $\odm$ that corresponds to opening
of $h_2 h_2$ final state. 

Constraints coming from the direct detection can be compared with results of \cite{Gross:2015cwa}. 
However, a detailed comparison is quite complicated, therefore we limit ourself to a conclusion that qualitatively
results obtained in \cite{Gross:2015cwa} for the Abelian case agree with those found here.
The results presented in the fig.~2 therein present similar behaviour to those of fig.~\ref{DM-N_xsec_2} in this work. 
It can be seen that for a given $\mtwo$, when $\mzp$ approaches $\mtwo/2$ the nucleon scattering cross section diminishes 
and LUX bounds can be easily satisfied, whereas for $\mzp$ between $(\mone+\mtwo)/2$ and $\mtwo$ the cross section is 
substantially larger (the points for $\mtwo<600$~GeV). Similarly the LUX bounds constrain the vicinity of $\mzp=80$~GeV.

\section{Summary}
\label{summary}
\setcounter{equation}{0}
In this paper we explored an extension of the Standard Model gauge symmetry by an
extra $\uone$ factor. The scalar sector of the model
consists of a standard $SU(2)_L$ doublet ($H$) and a complex scalar ($S$) charged under this
$\uone$ in order to provide a mass for the extra gauge boson. 
The possible spectrum of scalar states and mixings between them have been
discussed in detail. Adopting 2-loop RGE running we have required the vacuum 
stability (positivity) conditions to be valid up to the imposed cutoff scale ($Q=M_{Pl}$) of the model. 

We have compared the sensitivity of the running of the scalar quartic couplings and the stability of the vacuum 
on the loop order of the RGEs and found that the 2-loop RGEs make a significant effect on the 1-loop result, 
which could be improved further with higher order corrections. 

We found that an increase in $\lambda_H(m_t)$ by even a modest amount easily eliminates the 
electroweak vacuum stability problem up to the Planck scale, which is possible due to the 
additional degrees of freedom in this theory. When the Higgs portal $\kappa |S|^2 |H|^2$
is present (so $\kappa\neq 0$) then the initial value $\lamh(\mt)$ could be chosen larger than its Standard
Model value (where it is fixed by the Higgs mass measurement) in this way escaping the 
dangerous possibility of $\lamh(Q)$ becoming negative.  In addition the model discussed
here can have a stable vacuum even if $\lambda_H(m_t)$ is chosen {\it below} its Standard Model value, 
since the RGE beta function for $\lamh$ contains an extra {\it positive} contribution proportional to $ \kappa^2$
which can lift $\lamh$ efficiently enough while running up. For this mechanism to work
$\kappa(\mt)$ must be large enough.

The presence of an extra neutral scalar makes the phenomenology of the scalar sector 
more attractive and richer but still testable at the LHC and future colliders.
We have focused here on dark matter
aspects of the model. The parameter space of the models has been investigated
in detail and regions consistent with theoretical, collider and cosmological constraints
have been determined. In particular we have shown that the DM-nucleon
cross section could be consistent with the LUX and XENON100 limits, and also looked at why this is so. It has also been shown
that the anticipated XENON 1T limits on $\xdd$ may be satisfied within this model.
 
\section*{Acknowledgments}
BG and MD acknowledge partial support by the National Science Centre (Poland) research project, 
decision no DEC-2014/13/B/ST2/03969.
This work was partially supported by the Foundation for Polish Science International PhD Projects 
Programme co-financed by the EU European Regional Development Fund. This work has also been partially 
supported by National Science Centre (Poland) under research grant DEC-2012/04/A/ST2/00099.

\begin{appendix}

\section{Relic abundance}
\label{app:rel_abun}
In this appendix we present a detailed account of the tree-level calculation of the relic abundance in this model. 
This has been used to compare with the result generated with  \MICRO.

\subsection{Thermally averaged cross-section}
\label{averaged_xsec}
Thermally averaged cross-section is defined as the annihilation cross seciton $\sigma$ times M\o ller velocity $v$ averaged over the Boltzmann distribution
\begin{equation}
 \langle \sigma v \rangle = \frac{\int \sigma v \exp(-\sqrt{p^2_1+\mzp^2}/T)\exp(-\sqrt{p^2_2+\mzp^2}/T)d^3p_1d^3p_2}{\int \exp(-\sqrt{p^2_1+\mzp^2}/T)\exp(-\sqrt{p^2_2+\mzp^2}/T)d^3p_1d^3p_2}.
\end{equation}
In the nonrelativistic limit $m\gg T$ it can be approximated using the formula
\begin{equation}
 \langle \sigma |v|\rangle =\left[ \frac{\hat\sigma(s)}{4 M_{\zp}^2} + 
 \left(\frac{3}{2}\hat\sigma'(s)-\frac{\hat\sigma(s)}{4M_{\zp}^4}\right)\frac{1}{x} + O\left(\frac{1}{x^2}\right)\right]_{s=4M_{\zp}^2},
 \label{tav}
\end{equation}
where $x\equiv \frac{\mzp}{T}$ and $\hat\sigma(s)=2\sqrt{s(s-4\mzp^2)}\sigma(s)$ can be written as the sum of contributions coming from all possible final states
\begin{equation}
 \hat\sigma = \sum_f \hat\sigma_{\bar{f}f} + \hat\sigma_{W^+W^-} + \hat\sigma_{ZZ} + \hat\sigma_{h_1h_1} + \hat\sigma_{h_1h_2} + \hat\sigma_{h_2h_2}.
\end{equation}
Cross sections $\hat\sigma$ for fermions and vector bosons (presented in \ref{Zann}) can be easily expressed as a functions of $s$ only,
but for annihilations into scalars the amplitudes squared depend also on the Mandelstam variable $t$. Therefore respective $\hat\sigma(s)$ needs to be written as
\begin{equation}
 \hat{\sigma}_{h_i h_j}(s)=\frac{1}{9(1+\delta_{ij})}\beta(M_{h_i},M_{h_j})\int^{\pi}_0 d\theta \frac{\sin\theta}{2} f(s,t) d\theta, 
\end{equation}
where $f(s,t)$ is the respective sum of all amplitudes squared for given $i,j$ and $t$ can be expressed by $s$ and $\theta$ using
\begin{equation}
\begin{split}
 t=&M_{h_i}^2+\mzp^2- \sqrt{\frac{-2 s \left(M_{h_i}^2+M_{h_j}^2\right)+  \left(M_{h_i}^2-M_{h_j}^2\right)^2+s^2}{4}+s M_{h_i}^2}\\
   &+\frac{1}{2} \sqrt{\frac{s^2-4 \mzp^2 s}{s}} \sqrt{\frac{-2 s
   \left(M_{h_i}^2+M_{h_j}^2\right)+\left(M_{h_i}^2-M_{h_j}^2\right)^2+s^2}{s}}\cos\theta.	
\label{tform}
\end{split}
\end{equation}
In the nonrelativistic limit, using the formula ($\ref{tav}$), this tedious integration of $f(s,t)$ can be avoided by changing the order of the limit and the integral \cite{Roszkowski:1994tm}, then
\begin{equation}
 \hat\sigma_{h_ih_j}(s=4M_{\zp}^2) = \frac{1}{9(1+\delta_{ij})}\beta(M_{h_i},M_{h_j}) f(4M_{\zp}^2,t(s=4M_{\zp}^2)),
\end{equation}
where $t(s=4M_{\zp}^2)=(M_{h_i}^2+M_{h_j}^2)/2 - \mzp^2$; note that this expression does not depend on $\cos\theta$, therefore the integration is trivial and gives factor~$1$. Similarly for the derivative we have
\begin{equation}
\begin{split}
  \hat\sigma_{h_ih_j}'(s&=4M_{\zp}^2) = \frac{1}{9(1+\delta_{ij})}\beta(M_{h_i},M_{h_j}) \; \times \\
  &\left[\frac{1}{\beta}\frac{d\beta}{ds}f(s,t) + 
 \frac{\partial f(s,t)}{\partial s}  -\frac{1}{2}  \frac{\partial f(s,t)}{\partial t} \; + \right.\\
&+\left.\left((M^2_{h_i}-M^2_{h_j})^2+\frac{1}{12}(2M_{\zp}^2-M^2_{h_i}-M^2_{h_j})\right) \frac{\partial^2 f(s,t)}{\partial t^2}\right]_{s=4M_{\zp}^2,\;t=t(s=4M_{\zp}^2)}.
\end{split}
\end{equation}
The presence of the last term comes from the fact, that $t'(s)=a(s)+b(s)\cos\theta$, where $a(4M_{\zp}^2)=-1/2$ and 
$b(4M_{\zp}^2)=\infty$. On the other hand integral of $\cos\theta$ vanishes, therefore one needs to calculate the second derivative to determine the limit.
\subsection{Relevant vertices}
\label{Frules}
\begin{table}[h]
\begin{center}
\te
\begin{tabular}{|c|c|c|c|c|}
\hline
& &    & & \vspace{-0.5cm} 
\\
$V^{\zp}_i=i\frac{2M^2_{\zp}}{\vx} R_{2i}$
& $i\frac{2M^2_{Z}}{v}R_{1i}$
& $i\frac{2M^2_{W}}{v}R_{1i}$
& $i\frac{M_{F}}{v}R_{1i}$
& $V^{\zp}_{ij} = i\frac{2 M^2_{\zp}}{\vx^2}R_{2i}R_{2j}$
\\
& &    & & \vspace{-0.5cm} 
\\
\hline
& &    & & \vspace{-0.5cm} 
\\  
\begin{fmffile}{ZpZphi}
	        \begin{fmfgraph*}(60,50)
	            \fmfleft{i,j}
	            \fmfright{k}
	            \fmf{boson,label=$\zp$}{i,v}
		    \fmf{boson,label=$\zp$}{j,v}
		    \fmf{dashes,label=$h_i$}{v,k}
	        \end{fmfgraph*}
\end{fmffile} 
&
 \begin{fmffile}{ZZhi}
	        \begin{fmfgraph*}(60,50)
	            \fmfleft{i,j}
	            \fmfright{k}
	            \fmf{boson,label=$Z$}{i,v}
		    \fmf{boson,label=$Z$}{j,v}
		    \fmf{dashes,label=$h_i$}{v,k}
	        \end{fmfgraph*}
\end{fmffile}  & 
 \begin{fmffile}{WWhi}
	        \begin{fmfgraph*}(60,50)
	            \fmfleft{i,j}
	            \fmfright{k}
	            \fmf{boson,label=$W^+$}{i,v}
		    \fmf{boson,label=$W^-$}{j,v}
		    \fmf{dashes,label=$h_i$}{v,k}
	        \end{fmfgraph*}
\end{fmffile} &
 \begin{fmffile}{FFhi}
	        \begin{fmfgraph*}(60,50)
	            \fmfleft{i,j}
	            \fmfright{k}
	            \fmf{fermion,label=$\bar{f}$}{v,i}
		    \fmf{fermion,label=$f$}{j,v}
		    \fmf{dashes,label=$h_i$}{v,k}
	        \end{fmfgraph*}
\end{fmffile}&
\begin{fmffile}{hhZpZp}
	        \begin{fmfgraph*}(60,50)
	            \fmfleft{i,j}
	            \fmfright{k,l}
	            \fmf{boson,label=$\zp$}{i,v1}
		    \fmf{boson,label=$\zp$}{j,v1}
		    \fmf{dashes,label=$h_i$}{k,v1}
		    \fmf{dashes,label=$h_j$}{v1,l}
	        \end{fmfgraph*}
\end{fmffile}
\\\hline
\multicolumn{4}{|c|}{\hspace*{-2.5cm} \vbox{\vspace{-0.1cm}\begin{equation*}
\begin{split}
 V^h_{ijk}&= i[\kappa v (R_{1i} R_{2j} R_{2k} + R_{2i} R_{1j} R_{2k} + R_{2i} R_{2j} R_{1k}) \\ 
 &+ \kappa \vx (R_{2i} R_{1j} R_{1k} + R_{1i} R_{2j} R_{1k} + R_{1i} R_{1j} R_{2k})\\
 &+ 6\lambda v ( R_{1i} R_{1j} R_{1k} ) + 6 \lambda_s \vx (R_{2i} R_{2j} R_{2k})]
 \end{split}
\end{equation*}\vspace{-0.3cm}}\hspace*{-2.5cm}} & 
 \begin{fmffile}{hihjhk}
	        \begin{fmfgraph*}(60,50)
	            \fmfleft{i,j}
	            \fmfright{k}
	            \fmf{dashes,label=$h_i$}{i,v1}
		    \fmf{dashes,label=$h_j$}{j,v1}
		    \fmf{dashes,label=$h_k$}{k,v1}
	        \end{fmfgraph*}
\end{fmffile}
\\\hline
\end{tabular}
\end{center}
\end{table}
Scalar mixing matrix\begin{equation}
 R = \left( 
\begin{array}{cc}
\cos \alpha   &-\sin \alpha \\ 
\sin \alpha & \cos \alpha
\end{array} 
\right)
\end{equation}

\subsection{Useful formulas}
\label{formulas}

\begin{itemize}
 \item Lorentz invariant phase space
 \begin{equation}
 \begin{split}
  \beta(m_1,m_2)=&\frac{1}{8\pi}\sqrt{1-\frac{2(m_1^2+m_2^2)}{s}+\frac{(m_1^2-m_2^2)^2}{s^2}}\\
  \beta(m)=&\frac{1}{8\pi}\sqrt{1-\frac{4m^2}{s}}
 \end{split}
 \end{equation}
 \item Sum over polarizations of a vector boson
 \begin{equation}
  \Xi(m)=3-\frac{s}{m^2}+\frac{s^2}{4m^4}
 \end{equation}
 \item Sum over spins of a fermion
 \begin{equation}
    \Theta(m)=2s(1-\frac{4m^2}{s})
 \end{equation}
 \item Propagators
 \begin{equation}
 \begin{split}
  G_1(p)=&\frac{i}{p^2-M_{h_1}^2+i\Gamma_{h_1} M_{h_1}},\\
  G_2(p)=&\frac{i}{p^2-\mtwo^2+i\Gamma_{h_2} \mtwo},\\
  \Sigma^2=&\left|G_1(\sqrt{s})-G_2(\sqrt{s}) \right|^2=\frac{(\Gamma_1 M_{h_1}-\Gamma_2 M_{h_2})^2+(M_{h_1}^2-M_{h_2}^2)^2}{(\Gamma_1^2
   M_{h_1}^2+(s-M_{h_1})^2)((s-M_{h_2}^2)^2+\Gamma_2^2 M_{h_2}^2)},\\
  \Sigma_h = &i\left(V^{Z'}_1V^h_{1ij} G_1(\sqrt{s}) + V^h_{2ij} V^{Z'}_2 G_2(\sqrt{s}) \right)
 \end{split}
 \end{equation}

\end{itemize}

\subsection{$\zp$ annihilation cross section formulae}
\label{Zann}

\begin{flushleft}
\begin{tabular}{MM}
\hspace{-1cm}
\begin{fmffile}{ZZ1}
\begin{fmfgraph*}(120,60)
	            \fmfleft{i,j}
	            \fmfright{k,l}
	            \fmf{boson,label=$\zp$}{i,v1}
		    \fmf{boson,label=$\zp$}{j,v1}
		    \fmf{dashes,label=$h_1,,h_2$}{v1,v2}
		    \fmf{boson,label=$Z$}{v2,k}
		    \fmf{boson,label=$Z$}{v2,l}
	        \end{fmfgraph*}
\end{fmffile}
&
\parbox{10.8cm}{
\begin{equation}
 \hat{\sigma}_{ZZ}(s)=\frac{\beta(M_Z)}{18}  \Xi(\mzp)\Xi(M_Z) \left(\frac{4 M_Z \mzp}{v \vx}\right)^2 \cos^2\alpha\sin^2\alpha\Sigma^2
\end{equation}
}\\ \noalign{\smallskip}\noalign{\smallskip}\noalign{\smallskip}
\hspace{-1cm}
\begin{fmffile}{WW1}
	        \begin{fmfgraph*}(120,60)
	            \fmfleft{i,j}
	            \fmfright{k,l}
	            \fmf{boson,label=$\zp$}{i,v1}
		    \fmf{boson,label=$\zp$}{j,v1}
		    \fmf{dashes,label=$h_1,,h_2$}{v1,v2}
		    \fmf{boson,label=$W^+$}{v2,k}
		    \fmf{boson,label=$W^-$}{v2,l}
	        \end{fmfgraph*}
\end{fmffile} & 
\parbox{10.8cm}{
\begin{equation}
 \hat{\sigma}_{W^+W^-}(s)=\frac{\beta(M_W)}{9}  \Xi(\mzp)\Xi(M_W) \left(\frac{4 M_W \mzp}{v \vx}\right)^2 \cos^2\alpha\sin^2\alpha\Sigma^2
\end{equation}
}
\\\noalign{\smallskip}\noalign{\smallskip}\noalign{\smallskip}
\hspace{-1cm}
\begin{fmffile}{ff1}
	        \begin{fmfgraph*}(120,60)
	            \fmfleft{i,j}
	            \fmfright{k,l}
	            \fmf{boson,label=$\zp$}{i,v1}
		    \fmf{boson,label=$\zp$}{j,v1}
		    \fmf{dashes,label=$h_1,,h_2$}{v1,v2}
		    \fmf{fermion,label=$\bar{f}$}{k,v2}
		    \fmf{fermion,label=$f$}{v2,l}
	        \end{fmfgraph*}
\end{fmffile} & 
\parbox{10.8cm}{
\begin{equation}
 \hat{\sigma}_{\bar{f}f}(s)=\frac{\beta(M_F)}{9}\Theta(M_F)\Xi(\mzp)\left(\frac{2 M_F \mzp}{v \vx}\right)^2 \cos^2\alpha\sin^2\alpha\Sigma^2
\end{equation}
}
\end{tabular}
\end{flushleft}

\begin{center}
\begin{tabular}{llll}
\begin{fmffile}{hh1}
	        \begin{fmfgraph*}(60,60)
	            \fmfleft{i,j}
	            \fmfright{k,l}
	            \fmf{boson,label=$\zp$}{i,v1}
		    \fmf{boson,label=$\zp$}{j,v1}
		    \fmf{dashes,label=$h_i$}{k,v1}
		    \fmf{dashes,label=$h_j$}{v1,l}
	        \end{fmfgraph*}
\end{fmffile} & 
 
\begin{fmffile}{hh2}
	        \begin{fmfgraph*}(80,60)
	            \fmfleft{i,j}
	            \fmfright{k,l}
	            \fmf{boson,label=$\zp$}{i,v1}
		    \fmf{boson,label=$\zp$}{j,v1}
		    \fmf{dashes,label=$h_k$}{v1,v2}
		    \fmf{dashes,label=$h_i$}{k,v2}
		    \fmf{dashes,label=$h_j$}{v2,l}
	        \end{fmfgraph*}
\end{fmffile} & 
	        
\begin{fmffile}{hh3}
	        \begin{fmfgraph*}(80,60)
	            \fmfleft{i,j}
	            \fmfright{k,l}
	            \fmf{boson,label=$\zp$}{i,v1}
		    \fmf{boson,label=$\zp$}{j,v2}
		    \fmf{boson,label=$\zp$}{v1,v2}
		    \fmf{dashes,label=$h_i$}{k,v1}
		    \fmf{dashes,label=$h_j$}{v2,l}
	        \end{fmfgraph*} 
\end{fmffile} &
\begin{fmffile}{hh4}
	        \begin{fmfgraph*}(80,60)
		    \fmfleft{i,j}
	            \fmfright{k,l}
	            \fmf{boson,tension=0.6,label=$\zp$}{i,v1}
		    \fmf{boson,tension=0.6,label=$\zp$}{j,v2}
		    \fmf{boson,tension=-0.45,label=$\zp$}{v1,v2}
		    \fmf{dashes,tension=0.5,label=$h_j$,label.dist=12}{k,v2}
		    \fmf{dashes,tension=0.5,label=$h_i$,label.dist=12}{v1,l}
	        \end{fmfgraph*} 
\end{fmffile}
\end{tabular}
\end{center}
Squares of amplitudes and interference terms that contribute to the 4 diagrams contributing to the $h_ih_j$ ($i,j=1,2$)
final state shown above
are listed below where $f_k$ stands for the square of the $k$th amplitude while $f_{kl}$ for $2\Re(f_k f_l^\star)$ 
($k,l=1,2,3,4$).
{\footnotesize
\begin{flalign}
& f_1(s)=|V^{\zp}_{ij}|^2 \Xi(\mzp) &
\end{flalign}\begin{flalign}
&f_2(s)=|\Sigma_h|^2 \Xi(\mzp) &
\end{flalign}
\begin{flalign}
 &f_3(s,t)=|V^{\zp}_i V^{\zp}_j|^2 \frac{1}{(t-M_{\zp}^2)^2} \Big[2-\frac{2t}{M_{\zp}^2}+
 \frac{(-\mzp^2 + s/2)^2 + \frac{1}{2} (M_{h_j}^2 - \mzp^2 - t)^2 + t^2 + 
 \frac{1}{2} (-M_{h_i}^2 + \mzp^2 + t)^2}{16 M_{\zp}^4}+\nonumber\\
 &\frac{-(\frac{1}{4}) (M_{h_j}^2 - \mzp^2 - t)^2 t - 
 \frac{1}{2} (-\mzp^2 + s/2) (M_{h_j}^2 - \mzp^2 - t) (-M_{h_i}^2 + \mzp^2 + t) - 
 \frac{1}{4} t (-M_{h_i}^2 + \mzp^2 + t)^2}{M_{\zp}^6}\nonumber\\
 &+\frac{(M_{h_j}^2 - \mzp^2 - t)^2 (-M_{h_i}^2 + \mzp^2 + t)^2}{16M_{\zp}^8} \Big]
\end{flalign}
\begin{flalign}
 &f_4(s,t)=|V^{\zp}_i V^{\zp}_j|^2 \frac{1}{(M_{h_i}^2 + M_{h_j}^2 + \mzp^2 - s - t)^{2}} \Big[2-\frac{2 (M_{h_i}^2 + M_{h_j}^2 + 2 \mzp^2 - s - t)}{M_{\zp}^2}\nonumber\\
 &+\frac{(-\mzp^2 + s/2)^2 + (M_{h_i}^2 + M_{h_j}^2 + 2 \mzp^2 - s - t)^2 + 
 \frac{1}{2} (M_{h_i}^2 + 3 \mzp^2 - s - t)^2 + \frac{1}{2} (-M_{h_j}^2 - 3 \mzp^2 + s + t)^2}{M_{\zp}^4}-\nonumber\\
 &\frac{\frac{1}{2} (M_{h_i}^2 + M_{h_j}^2 + 2 \mzp^2 - s - t) (M_{h_i}^2 + 3 \mzp^2 - s - t)^2 +
  ( s/2 - \mzp^2) (M_{h_i}^2 + 3 \mzp^2 - s - t) (-M_{h_j}^2 - 3 \mzp^2 + s + 
 t)}{2M_{\zp}^6}\nonumber\\
 &+\frac{ \frac{1}{2}(M_{h_i}^2 + M_{h_j}^2 + 2 \mzp^2 - s - t) (-M_{h_j}^2 - 3 \mzp^2 + s + 
 t)^2}{2M_{\zp}^6} +\frac{(M_{h_i}^2 + 3 \mzp^2 - s - t)^2 (-M_{h_j}^2 - 3 \mzp^2 + s + t)^2}{16M_{\zp}^8} \Big]
 \end{flalign}
\begin{flalign}
 &f_{12}(s)=2 |V^{\zp}_{ij}|\Xi(\mzp)\Re(\Sigma_h)&
\end{flalign}
\begin{flalign}
 &f_{13}(s,t)=2 |V^{\zp}_{ij} V^{\zp}_{i}V^{\zp}_{j}| \frac{1}{t-M_{\zp}^2} 
 \Big [2 - \frac{2t}{M_{\zp}^2}  + \frac{(-\mzp^2 + s/2)^2 + \frac{1}{4} (M_{h_j}^2 - \mzp^2 - t)^2 + 
 \frac{1}{4} (-M_{h_i}^2 + \mzp^2 + t)^2}{M_{\zp}^4}\nonumber\\
 &+ \frac{(-\mzp^2 + s/2) (M_{h_j}^2 - \mzp^2 - t) (-M_{h_i}^2 + \mzp^2 + t)}{4M_{\zp}^6}\Big]
\end{flalign}
\begin{flalign}
 &f_{14}(s,t)=2 |V^{\zp}_{ij} V^{\zp}_{i}V^{\zp}_{j}| \frac{1}{M_{h_i}^2 + M_{h_j}^2 + \mzp^2 - s - t} 
 \Big [2 - \frac{M_{h_i}^2 + M_{h_j}^2 + 2 \mzp^2 - s - t}{M_{\zp}^2}\nonumber\\
 &- \frac{(-\mzp^2 + s/2) (M_{h_i}^2 + 3 \mzp^2 - s - t) (-M_{h_j}^2 - 3 \mzp^2 + s + t)}{M_{\zp}^4}\nonumber\\
 &+ \frac{(-\mzp^2 + s/2)^2 + \frac{1}{4} (M_{h_i}^2 + 3 \mzp^2 - s - t)^2 + 
 \frac{1}{4} (-M_{h_j}^2 - 3 \mzp^2 + s + t)^2}{4M_{\zp}^6}\Big]&
\end{flalign}
\begin{flalign}
 &f_{23}(s,t)=2 \Re(\Sigma_h) |V^{\zp}_{i}V^{\zp}_{j}| \frac{1}{t-\mzp^2} 
  \Big [2 - \frac{t}{M_{\zp}^2}  + \frac{(-\mzp^2 + s/2)^2 + \frac{1}{4} (M_{h_j}^2 - \mzp^2 - t)^2 + 
 \frac{1}{4} (-M_{h_i}^2 + \mzp^2 + t)^2}{M_{\zp}^4}\nonumber\\
 &+ \frac{(-\mzp^2 + s/2) (M_{h_j}^2 - \mzp^2 - t) (-M_{h_i}^2 + \mzp^2 + t)}{4M_{\zp}^6}\Big]
\end{flalign}
\begin{flalign}
 &f_{24}(s,t)=2 \Re(\Sigma_h) |V^{\zp}_{i}V^{\zp}_{j}|  \frac{1}{M_{h_i}^2 + M_{h_j}^2 + \mzp^2 - s - t} 
 \Big [2 - \frac{M_{h_i}^2 + M_{h_j}^2 + 2 \mzp^2 - s - t}{M_{\zp}^2}  \nonumber\\
 &- \frac{(-\mzp^2 + s/2) (M_{h_i}^2 + 3 \mzp^2 - s - t) (-M_{h_j}^2 - 3 \mzp^2 + s + t)}{M_{\zp}^4}\nonumber\\
 &+ \frac{(-\mzp^2 + s/2)^2 + \frac{1}{4} (M_{h_i}^2 + 3 \mzp^2 - s - t)^2 + 
 \frac{1}{4} (-M_{h_j}^2 - 3 \mzp^2 + s + t)^2}{4M_{\zp}^6}\Big]&
\end{flalign}
\begin{flalign}
  &f_{34}(s,t)=2|V^{\zp}_{i}V^{\zp}_{j}|^2 \frac{1}{(M_{h_i}^2 + M_{h_j}^2 + \mzp^2 - s - t)(t-M_{\zp}^2)} 
 \Big [2 + \frac{-M_{h_i}^2 - M_{h_j}^2 - 2 \mzp^2 + s}{M_{\zp}^2} \nonumber\\
 &- \frac{4(-\mzp^2 + s/2)^2 + (M_{h_j}^2 - \mzp^2 - t)^2 + 
  (M_{h_i}^2 + 3 \mzp^2 - s - t)^2 + (-M_{h_i}^2 + \mzp^2 + t)^2 + 
 (-M_{h_j}^2 - 3 \mzp^2 + s + t)^2}{4 M_{\zp}^4}\nonumber\\
 &- \frac{(-\mzp^2 + s/2) (M_{h_j}^2 - \mzp^2 - t) (-M_{h_i}^2 + \mzp^2 + t)+(-\mzp^2 + s/2) (M_{h_i}^2 + 3 \mzp^2 - s - t) (-M_{h_j}^2 - 3 \mzp^2 + s + t)}{16M_{\zp}^6}\nonumber\\
 &\frac{4(-\mzp^2 + s/2)^2 + (M_{h_j}^2 - \mzp^2 - t)^2 + 
 (M_{h_i}^2 + 3 \mzp^2 - s - t)^2 + \frac{1}{4} (-M_{h_i}^2 + \mzp^2 + t)^2 + 
 (-M_{h_j}^2 - 3 \mzp^2 + s + t)^2}{4 M_{\zp}^8} \Big]
\end{flalign}}


\section{Model files for \SARAH}
\label{app:modelfiles}
In this appendix are included the model files used for \SARAH \cite{Staub:2009bi,Staub:2010jh,Staub:2012pb,Staub:2013tta}, to study the model
discussed in this paper.   Currently \SARAH does not implement a $Z_2$ symmetry for the imaginary scalar component to give $S\rightarrow S^*$, nor for the $Z_2$ for the $\uone$ gauge field.  Instead the kinetic mixing couplings have been set to vanish,  $g_{1,x}=g_{x,1}\equiv0$, to preserve this symmetry.


\subsection{\tt SMVDM.m}
\lstset{basicstyle=\scriptsize,
frame=shadowbox}
\begin{lstlisting}
(*-------------------------------------------*)
(*   Particle Content                        *)
(*-------------------------------------------*)

(* Gauge Superfields *)

Gauge[[1]]={B,   U[1], hypercharge, g1, False};
Gauge[[2]]={WB, SU[2], left,        g2,  True};
Gauge[[3]]={G,  SU[3], color,       g3, False};
Gauge[[4]]={X,   U[1], darkcharge,  gx, False};

(* Chiral Superfields *)

FermionFields[[1]] = {q, 3, {uL, dL},     1/6, 2,  3, 0};  
FermionFields[[2]] = {l, 3, {vL, eL},    -1/2, 2,  1, 0};
FermionFields[[3]] = {d, 3, conj[dR],     1/3, 1, -3, 0};
FermionFields[[4]] = {u, 3, conj[uR],    -2/3, 1, -3, 0};
FermionFields[[5]] = {e, 3, conj[eR],       1, 1,  1, 0};

(* FermionFields[[6]] = {v, 3, conj[vR],   0, 1,  1, 0}; *)

ScalarFields[[1]] =  {H, 1, {Hp, H0}, 1/2, 2, 1, 0};
ScalarFields[[2]] =  {S, 1,       hs,   0, 1, 1, 1};

(*----------------------------------------------*)
(*   DEFINITION                                 *)
(*----------------------------------------------*)

NameOfStates={GaugeES, EWSB};

(* ----- Before EWSB ----- *)

DEFINITION[GaugeES][DiracSpinors]={
  Fd1 -> {dL, 0},
  Fd2 -> {0, dR},
  Fu1 -> {uL, 0},
  Fu2 -> {0, uR},
  Fe1 -> {eL, 0},
  Fe2 -> {0, eR},
   Fv -> {vL, 0}};

DEFINITION[GaugeES][Additional]= {
	{  LagHC, {Overwrite->True, AddHC->True}},
	{LagNoHC, {Overwrite->True, AddHC->False}}};

LagNoHC = (mu2 conj[H].H - lambdaH conj[H].H.conj[H].H + nu2 conj[S].S
 		- lambdaS conj[S].S.conj[S].S - L3 conj[S].S.conj[H].H);

LagHC =  -( Yd conj[H].d.q + Ye conj[H].e.l + Yu H.u.q );

(* Gauge Sector *)

DEFINITION[EWSB][GaugeSector] =
{ 
  {{VB,VWB[3]},{VP,VZ},ZZ},
  {{VX},  {VZp},ZZp},
  {{VWB[1],VWB[2]},{VWp,conj[VWp]},ZW}};     	

(* ----- VEVs ---- *)
 
DEFINITION[EWSB][VEVs]= 
{    {H0, {vSM, 1/Sqrt[2]}, {sigmaH, \[ImaginaryI]/Sqrt[2]},{phiH, 1/Sqrt[2]}},
     {hs,  {vX, 1/Sqrt[2]}, {sigmaS, \[ImaginaryI]/Sqrt[2]},{phiS, 1/Sqrt[2]}}};

DEFINITION[EWSB][MatterSector]=   
    {(*     {{vL,conj[vR]}, {VL,ZM}}	*)
     {{phiH,phiS},{hh,ZH}},
     {{sigmaH,sigmaS},{Ah,ZA}},
     {{{dL}, {conj[dR]}}, {{DL,Vd}, {DR,Ud}}},
     {{{uL}, {conj[uR]}}, {{UL,Vu}, {UR,Uu}}},
     {{{eL}, {conj[eR]}}, {{EL,Ve}, {ER,Ue}}}}; 

(*------------------------------------------------------*)
(* Dirac-Spinors                                        *)
(*------------------------------------------------------*)

DEFINITION[EWSB][DiracSpinors]={
 Fd ->{  DL, conj[DR]},
 Fe ->{  EL, conj[ER]},
 Fu ->{  UL, conj[UR]},
 Fv ->{  vL, 0}};

DEFINITION[EWSB][GaugeES]={
 Fd1 ->{  FdL, 0},
 Fd2 ->{  0, FdR},
 Fu1 ->{  Fu1, 0},
 Fu2 ->{  0, Fu2},
 Fe1 ->{  Fe1, 0},
 Fe2 ->{  0, Fe2}};

\end{lstlisting}

\subsection{\tt SPheno.m}
\begin{lstlisting}
(*----------------------------------------------*)
(*   MINPAR                                     *)
(*----------------------------------------------*)

OnlyLowEnergySPheno = True;

MINPAR={{1,lambdaHINPUT},
        {2,L3INPUT},
        {3,lambdaSINPUT},
        {10, gxINPUT},
        {20, vXInput} };

ParametersToSolveTadpoles = {nu2,mu2};

RealParameters = {vSM,vX};

BoundaryLowScaleInput={
 {gx,gxINPUT},
 {g1x,0},
 {gx1,0},
 {lambdaH, lambdaHINPUT},
 {lambdaS, lambdaSINPUT},
 {L3, L3INPUT},
 {vX, vXInput}};

BoundaryLowScaleInput={
 {vSM,Sqrt[4 mz2/(g1^2+g2^2)]}};

ListDecayParticles = {Fu,Fe,Fd,hh,VZp,Hp};
ListDecayParticles3B = {{Fu,"Fu.f90"},{Fe,"Fe.f90"},{Fd,"Fd.f90"}};

\end{lstlisting}

\end{appendix}

\bibliography{SMVDM}

\providecommand{\href}[2]{#2}\begingroup\raggedright\begin{thebibliography}{10}

\bibitem{Aad:2012tfa}
{\bf ATLAS} Collaboration, G.~Aad et~al., {\it {Observation of a new particle
  in the search for the Standard Model Higgs boson with the ATLAS detector at
  the LHC}},  {\em Phys.Lett.} {\bf B716} (2012) 1--29,
  [\href{http://xxx.lanl.gov/abs/1207.7214}{{\tt arXiv:1207.7214}}].

\bibitem{Chatrchyan:2012ufa}
{\bf CMS} Collaboration, S.~Chatrchyan et~al., {\it {Observation of a new boson
  at a mass of 125 GeV with the CMS experiment at the LHC}},  {\em Phys.Lett.}
  {\bf B716} (2012) 30--61, [\href{http://xxx.lanl.gov/abs/1207.7235}{{\tt
  arXiv:1207.7235}}].

\bibitem{Cabibbo:1979ay}
N.~Cabibbo, L.~Maiani, G.~Parisi, and R.~Petronzio, {\it {Bounds on the
  Fermions and Higgs Boson Masses in Grand Unified Theories}},  {\em
  Nucl.Phys.} {\bf B158} (1979) 295--305.

\bibitem{Hung:1979dn}
P.~Q. Hung, {\it {Vacuum Instability and New Constraints on Fermion Masses}},
  {\em Phys.Rev.Lett.} {\bf 42} (1979) 873.

\bibitem{Lindner:1985uk}
M.~Lindner, {\it {Implications of Triviality for the Standard Model}},  {\em
  Z.Phys.} {\bf C31} (1986) 295.

\bibitem{Lindner:1988ww}
M.~Lindner, M.~Sher, and H.~W. Zaglauer, {\it {Probing Vacuum Stability Bounds
  at the Fermilab Collider}},  {\em Phys.Lett.} {\bf B228} (1989) 139.

\bibitem{Arnold:1989cb}
P.~B. Arnold, {\it {Can the Electroweak Vacuum Be Unstable?}},  {\em Phys.Rev.}
  {\bf D40} (1989) 613.

\bibitem{Sher:1988mj}
M.~Sher, {\it {Electroweak Higgs Potentials and Vacuum Stability}},  {\em
  Phys.Rept.} {\bf 179} (1989) 273--418.

\bibitem{Sher:1993mf}
M.~Sher, {\it {Precise vacuum stability bound in the standard model}},  {\em
  Phys.Lett.} {\bf B317} (1993) 159--163,
  [\href{http://xxx.lanl.gov/abs/hep-ph/9307342}{{\tt hep-ph/9307342}}].

\bibitem{Schrempp:1996fb}
B.~Schrempp and M.~Wimmer, {\it {Top quark and Higgs boson masses: Interplay
  between infrared and ultraviolet physics}},  {\em Prog.Part.Nucl.Phys.} {\bf
  37} (1996) 1--90, [\href{http://xxx.lanl.gov/abs/hep-ph/9606386}{{\tt
  hep-ph/9606386}}].

\bibitem{Bertone:2004pz}
G.~Bertone, D.~Hooper, and J.~Silk, {\it {Particle dark matter: Evidence,
  candidates and constraints}},  {\em Phys.Rept.} {\bf 405} (2005) 279--390,
  [\href{http://xxx.lanl.gov/abs/hep-ph/0404175}{{\tt hep-ph/0404175}}].

\bibitem{Lebedev:2011iq}
O.~Lebedev, H.~M. Lee, and Y.~Mambrini, {\it {Vector Higgs-portal dark matter
  and the invisible Higgs}},  {\em Phys.Lett.} {\bf B707} (2012) 570--576,
  [\href{http://xxx.lanl.gov/abs/1111.4482}{{\tt arXiv:1111.4482}}].

\bibitem{Farzan:2012hh}
Y.~Farzan and A.~R. Akbarieh, {\it {VDM: A model for Vector Dark Matter}},
  {\em JCAP} {\bf 1210} (2012) 026,
  [\href{http://xxx.lanl.gov/abs/1207.4272}{{\tt arXiv:1207.4272}}].

\bibitem{Baek:2012se}
S.~Baek, P.~Ko, W.-I. Park, and E.~Senaha, {\it {Higgs Portal Vector Dark
  Matter : Revisited}},  {\em JHEP} {\bf 1305} (2013) 036,
  [\href{http://xxx.lanl.gov/abs/1212.2131}{{\tt arXiv:1212.2131}}].

\bibitem{Baek:2014jga}
S.~Baek, P.~Ko, and W.-I. Park, {\it {Invisible Higgs Decay Width vs. Dark
  Matter Direct Detection Cross Section in Higgs Portal Dark Matter Models}},
  {\em Phys.Rev.} {\bf D90} (2014) 055014,
  [\href{http://xxx.lanl.gov/abs/1405.3530}{{\tt arXiv:1405.3530}}].

\bibitem{Hambye:2008bq}
T.~Hambye, {\it {Hidden vector dark matter}},  {\em JHEP} {\bf 0901} (2009)
  028, [\href{http://xxx.lanl.gov/abs/0811.0172}{{\tt arXiv:0811.0172}}].

\bibitem{Gross:2015cwa}
C.~Gross, O.~Lebedev, and Y.~Mambrini, {\it {Non-Abelian gauge fields as dark
  matter}},  \href{http://xxx.lanl.gov/abs/1505.0748}{{\tt arXiv:1505.0748}}.

\bibitem{DiChiara:2015bua}
S.~Di~Chiara and K.~Tuominen, {\it {A minimal model for ${\rm SU}(N)$ vector
  dark matter}},  \href{http://xxx.lanl.gov/abs/1506.0328}{{\tt
  arXiv:1506.0328}}.

\bibitem{Yu:2014pra}
J.-H. Yu, {\it {Vector Fermion-Portal Dark Matter: Direct Detection and
  Galactic Center Gamma-Ray Excess}},  {\em Phys.Rev.} {\bf D90} (2014) 095010,
  [\href{http://xxx.lanl.gov/abs/1409.3227}{{\tt arXiv:1409.3227}}].

\bibitem{Ghorbani:2015baa}
K.~Ghorbani and H.~Ghorbani, {\it {Two-portal Dark Matter}},  {\em Phys.Rev.}
  {\bf D91} (2015) 123541, [\href{http://xxx.lanl.gov/abs/1504.0361}{{\tt
  arXiv:1504.0361}}].

\bibitem{Staub:2009bi}
F.~Staub, {\it {From Superpotential to Model Files for FeynArts and
  CalcHep/CompHep}},  {\em Comput.Phys.Commun.} {\bf 181} (2010) 1077--1086,
  [\href{http://xxx.lanl.gov/abs/0909.2863}{{\tt arXiv:0909.2863}}].

\bibitem{Staub:2010jh}
F.~Staub, {\it {Automatic Calculation of supersymmetric Renormalization Group
  Equations and Self Energies}},  {\em Comput.Phys.Commun.} {\bf 182} (2011)
  808--833, [\href{http://xxx.lanl.gov/abs/1002.0840}{{\tt arXiv:1002.0840}}].

\bibitem{Staub:2012pb}
F.~Staub, {\it {SARAH 3.2: Dirac Gauginos, UFO output, and more}},  {\em
  Computer Physics Communications} {\bf 184} (2013) pp. 1792--1809,
  [\href{http://xxx.lanl.gov/abs/1207.0906}{{\tt arXiv:1207.0906}}].

\bibitem{Staub:2013tta}
F.~Staub, {\it {SARAH 4: A tool for (not only SUSY) model builders}},  {\em
  Comput.Phys.Commun.} {\bf 185} (2014) 1773--1790,
  [\href{http://xxx.lanl.gov/abs/1309.7223}{{\tt arXiv:1309.7223}}].

\bibitem{Lyonnet:2013dna}
F.~Lyonnet, I.~Schienbein, F.~Staub, and A.~Wingerter, {\it {PyR@TE:
  Renormalization Group Equations for General Gauge Theories}},  {\em
  Comput.Phys.Commun.} {\bf 185} (2014) 1130--1152,
  [\href{http://xxx.lanl.gov/abs/1309.7030}{{\tt arXiv:1309.7030}}].

\bibitem{Porod:2011nf}
W.~Porod and F.~Staub, {\it {SPheno 3.1: Extensions including flavour,
  CP-phases and models beyond the MSSM}},  {\em Comput.Phys.Commun.} {\bf 183}
  (2012) 2458--2469, [\href{http://xxx.lanl.gov/abs/1104.1573}{{\tt
  arXiv:1104.1573}}].

\bibitem{Porod:2003um}
W.~Porod, {\it {SPheno, a program for calculating supersymmetric spectra, SUSY
  particle decays and SUSY particle production at e+ e- colliders}},  {\em
  Comput.Phys.Commun.} {\bf 153} (2003) 275--315,
  [\href{http://xxx.lanl.gov/abs/hep-ph/0301101}{{\tt hep-ph/0301101}}].

\bibitem{Belanger:2001fz}
G.~Belanger, F.~Boudjema, A.~Pukhov, and A.~Semenov, {\it {MicrOMEGAs: A
  Program for calculating the relic density in the MSSM}},  {\em
  Comput.Phys.Commun.} {\bf 149} (2002) 103--120,
  [\href{http://xxx.lanl.gov/abs/hep-ph/0112278}{{\tt hep-ph/0112278}}].

\bibitem{Belanger:2004yn}
G.~Belanger, F.~Boudjema, A.~Pukhov, and A.~Semenov, {\it {micrOMEGAs: Version
  1.3}},  {\em Comput.Phys.Commun.} {\bf 174} (2006) 577--604,
  [\href{http://xxx.lanl.gov/abs/hep-ph/0405253}{{\tt hep-ph/0405253}}].

\bibitem{Belanger:2006is}
G.~Belanger, F.~Boudjema, A.~Pukhov, and A.~Semenov, {\it {MicrOMEGAs 2.0: A
  Program to calculate the relic density of dark matter in a generic model}},
  {\em Comput.Phys.Commun.} {\bf 176} (2007) 367--382,
  [\href{http://xxx.lanl.gov/abs/hep-ph/0607059}{{\tt hep-ph/0607059}}].

\bibitem{Belanger:2013oya}
G.~Belanger, F.~Boudjema, A.~Pukhov, and A.~Semenov, {\it {micrOMEGAs3: A
  program for calculating dark matter observables}},  {\em Comput.Phys.Commun.}
  {\bf 185} (2014) 960--985, [\href{http://xxx.lanl.gov/abs/1305.0237}{{\tt
  arXiv:1305.0237}}].

\bibitem{Akerib:2013tjd}
{\bf LUX Collaboration} Collaboration, D.~Akerib et~al., {\it {First results
  from the LUX dark matter experiment at the Sanford Underground Research
  Facility}},  {\em Phys.Rev.Lett.} {\bf 112} (2014) 091303,
  [\href{http://xxx.lanl.gov/abs/1310.8214}{{\tt arXiv:1310.8214}}].

\bibitem{Aprile:2012nq}
{\bf XENON100 Collaboration} Collaboration, E.~Aprile et~al., {\it {Dark Matter
  Results from 225 Live Days of XENON100 Data}},  {\em Phys.Rev.Lett.} {\bf
  109} (2012) 181301, [\href{http://xxx.lanl.gov/abs/1207.5988}{{\tt
  arXiv:1207.5988}}].

\bibitem{Buttazzo:2013uya}
D.~Buttazzo, G.~Degrassi, P.~P. Giardino, G.~F. Giudice, F.~Sala, et~al., {\it
  {Investigating the near-criticality of the Higgs boson}},  {\em JHEP} {\bf
  1312} (2013) 089, [\href{http://xxx.lanl.gov/abs/1307.3536}{{\tt
  arXiv:1307.3536}}].

\bibitem{DiChiara:2014wha}
S.~Di~Chiara, V.~Keus, and O.~Lebedev, {\it {Stabilizing the Higgs potential
  with a Z$'$}},  {\em Phys.Lett.} {\bf B744} (2015) 59--66,
  [\href{http://xxx.lanl.gov/abs/1412.7036}{{\tt arXiv:1412.7036}}].

\bibitem{Gonderinger:2012rd}
M.~Gonderinger, H.~Lim, and M.~J. Ramsey-Musolf, {\it {Complex Scalar Singlet
  Dark Matter: Vacuum Stability and Phenomenology}},  {\em Phys.Rev.} {\bf D86}
  (2012) 043511, [\href{http://xxx.lanl.gov/abs/1202.1316}{{\tt
  arXiv:1202.1316}}].

\bibitem{Alanne:2014bra}
T.~Alanne, K.~Tuominen, and V.~Vaskonen, {\it {Strong phase transition, dark
  matter and vacuum stability from simple hidden sectors}},  {\em Nucl.Phys.}
  {\bf B889} (2014) 692--711, [\href{http://xxx.lanl.gov/abs/1407.0688}{{\tt
  arXiv:1407.0688}}].

\bibitem{Khan:2014kba}
N.~Khan and S.~Rakshit, {\it {A study of electroweak vacuum metastablity with a
  singlet scalar dark matter}},  \href{http://xxx.lanl.gov/abs/1407.6015}{{\tt
  arXiv:1407.6015}}.

\bibitem{Martin-Lozano:2015dja}
V.~Martin-Lozano, J.~M. Moreno, and C.~B. Park, {\it {Resonant Higgs boson pair
  production in the $hh\rightarrow b\bar{b} \; WW \rightarrow b\bar{b} \ell^+
  \nu \ell^- \bar\nu$ decay channel}},
  \href{http://xxx.lanl.gov/abs/1501.0379}{{\tt arXiv:1501.0379}}.

\bibitem{Robens:2015gla}
T.~Robens and T.~Stefaniak, {\it {Status of the Higgs Singlet Extension of the
  Standard Model after LHC Run 1}},  {\em Eur.Phys.J.} {\bf C75} (2015), no.~3
  104, [\href{http://xxx.lanl.gov/abs/1501.0223}{{\tt arXiv:1501.0223}}].

\bibitem{Falkowski:2015iwa}
A.~Falkowski, C.~Gross, and O.~Lebedev, {\it {A second Higgs from the Higgs
  portal}},  {\em JHEP} {\bf 1505} (2015) 057,
  [\href{http://xxx.lanl.gov/abs/1502.01361}{{\tt arXiv:1502.01361}}].

\bibitem{Aad:2015uga}
{\bf ATLAS} Collaboration, G.~Aad et~al., {\it {Search for invisible decays of
  the Higgs boson produced in association with a hadronically decaying vector
  boson in $pp$ collisions at $\sqrt{s}$ = 8 TeV with the ATLAS detector}},
  \href{http://xxx.lanl.gov/abs/1504.0432}{{\tt arXiv:1504.0432}}.

\bibitem{CMS:2015dia}
{\bf CMS} Collaboration, C.~Collaboration, {\it {Search for invisible decays of
  Higgs bosons in the vector boson fusion production mode}}, .

\bibitem{Chatrchyan:2014tja}
{\bf CMS} Collaboration, S.~Chatrchyan et~al., {\it {Search for invisible
  decays of Higgs bosons in the vector boson fusion and associated ZH
  production modes}},  {\em Eur.Phys.J.} {\bf C74} (2014) 2980,
  [\href{http://xxx.lanl.gov/abs/1404.1344}{{\tt arXiv:1404.1344}}].

\bibitem{Schael:2006cr}
{\bf ALEPH Collaboration, DELPHI Collaboration, L3 Collaboration, OPAL
  Collaboration, LEP Working Group for Higgs Boson Searches} Collaboration,
  S.~Schael et~al., {\it {Search for neutral MSSM Higgs bosons at LEP}},  {\em
  Eur.Phys.J.} {\bf C47} (2006) 547--587,
  [\href{http://xxx.lanl.gov/abs/hep-ex/0602042}{{\tt hep-ex/0602042}}].

\bibitem{ATLAS:Moriond_2015}
T.~A. collaboration, {\it {Measurements of the Higgs boson production and decay
  rates and coupling strengths using pp collision data at √s = 7 and 8 TeV in
  the ATLAS experiment}}, .

\bibitem{Khachatryan:2014jba}
{\bf CMS} Collaboration, V.~Khachatryan et~al., {\it {Precise determination of
  the mass of the Higgs boson and tests of compatibility of its couplings with
  the standard model predictions using proton collisions at 7 and 8 $\,\text
  {TeV}$}},  {\em Eur.Phys.J.} {\bf C75} (2015), no.~5 212,
  [\href{http://xxx.lanl.gov/abs/1412.8662}{{\tt arXiv:1412.8662}}].

\bibitem{Peskin:1991sw}
M.~E. Peskin and T.~Takeuchi, {\it {Estimation of oblique electroweak
  corrections}},  {\em Phys.Rev.} {\bf D46} (1992) 381--409.

\bibitem{Ciuchini:2014dea}
M.~Ciuchini, E.~Franco, S.~Mishima, M.~Pierini, L.~Reina, et~al., {\it {Update
  of the electroweak precision fit, interplay with Higgs-boson signal strengths
  and model-independent constraints on new physics}},
  \href{http://xxx.lanl.gov/abs/1410.6940}{{\tt arXiv:1410.6940}}.

\bibitem{Kolb:1990vq}
E.~W. Kolb and M.~S. Turner, {\it {The Early Universe}},  {\em Front.Phys.}
  {\bf 69} (1990) 1--547.

\bibitem{Ade:2015xua}
{\bf Planck} Collaboration, P.~Ade et~al., {\it {Planck 2015 results. XIII.
  Cosmological parameters}},  \href{http://xxx.lanl.gov/abs/1502.0158}{{\tt
  arXiv:1502.0158}}.

\bibitem{Cheng2014}
H.-Y. Cheng, {\it Scalar and pseudoscalar higgs couplings with nucleons},  {\em
  Nuclear Physics B - Proceedings Supplements} {\bf 246--247} (2014), no.~0 109
  -- 115. Proceedings of the 9th International Symposium on Cosmology and
  Particle Astrophysics Proceedings of the 9th International Symposium on
  Cosmology and Particle Astrophysics.

\bibitem{Roszkowski:1994tm}
L.~Roszkowski, {\it {A Simple way of calculating cosmological relic density}},
  {\em Phys.Rev.} {\bf D50} (1994) 4842--4845,
  [\href{http://xxx.lanl.gov/abs/hep-ph/9404227}{{\tt hep-ph/9404227}}].

\end{thebibliography}\endgroup
\bibliographystyle{JHEP}

\end{document}